\documentclass[]{aa}  

\usepackage{comment}
\usepackage{graphicx}
\usepackage{txfonts}
\usepackage{siunitx}
\usepackage{multirow}
\usepackage{amsmath, amssymb}
\usepackage{subfig}
\usepackage{ulem}

\newcommand\m[1]{%
\renewcommand\arraystretch{1}%
\begin{tabular}{@{}l@{}}#1\end{tabular}}

\newcolumntype{Q}{>{\tiny}c}
\begin{document}

   \title{Correlation between accretion rate and free-free emission in protoplanetary disks}
   \subtitle{A multiwavelength analysis of central mm/cm emission in transition disks}

   \author{ A. A. Rota
          \inst{\ref{leiden}},
          J. D. Meijerhof \inst{\ref{leiden}},
          N. van der Marel \inst{\ref{leiden}},
          L. Francis \inst{\ref{leiden}},
          F. F. S. van der Tak \inst{\ref{sron},\ref{kapteyn}},
          A. D. Sellek \inst{\ref{leiden}}
        }

    \institute{Leiden Observatory, Leiden University, P.O. Box 9513, 2300 RA Leiden, The Netherlands\label{leiden}\\ \email{rota@strw.leidenuniv.nl}
   \and
   SRON Netherlands Institute for Space Research, Landleven 12, NL-9747AD Groningen, The Netherlands\label{sron}
    \and
    Kapteyn Astronomical Institute, University of Groningen, NL-9747AD Groningen, The Netherlands\label{kapteyn}
              }
        
\titlerunning{A correlation between accretion rate and free-free emission in protoplanetary disks}
\authorrunning{Rota et al.}

   \date{Received -; accepted -}

 
  \abstract
{The inner regions of protoplanetary disks are  believed to be the primary locations of planet formation and the processes that influence the global evolution of the disk, such as magnetohydrodynamic winds and photoevaporation.
Transition disks with large inner dust cavities are ideal targets for studying the inner regions (of tens of au) of disks, as this is where the central emission can be fully disentangled from the outer disk emission.
}
{We present a homogeneous multiwavelength analysis of the continuum emission in a sample of 11 transition disks. We investigate the nature of the central emission close to the star, distinguishing between thermal dust and free-free emission.
}
 {We combined spatially resolved measurements of continuum emission from the archival Atacama Large Millimeter/Submillimeter Array data with centimeter-wave (cm-wave) observations from the literature to study the spectral indices of the inner and outer disks separately.}
 {While the emission from the outer disks is consistent with thermal dust emission, 10 out of 11 of the spectral indices estimated for the central emission close to the star suggest that this emission is free-free emission that is likely associated with an ionized jet or a disk wind. We found no correlation between the free-free luminosity and the accretion luminosity or the X-ray luminosity and this argues against an explanation based on a potential photoevaporative wind. A sub-linear correlation between the ionized mass loss rate and the accretion rate onto the star was observed, suggesting the origin is drawn from the ionized jet.}
 { The relative lack of millimeter-dust (mm-dust) grains in the majority of inner disks in transition disks indicates that either such dust grains have drifted quickly towards the central star, that grain growth is less efficient in the inner disk, or that grains  rapidly grow to planetesimal sizes in the inner disk. The observed correlation between the ionized mass loss rate and the accretion rate suggests the outflow is strictly connected to stellar accretion and  that accretion in these disks is driven by a jet. }


   \maketitle
%

\section{Introduction}\label{intro}

Protoplanetary disks surrounding young stars are the cradles where planets are formed. 
High-resolution and high-sensitivity surveys of protoplanetary disks conducted with the Atacama Large Millimeter/submillimeter Array (ALMA) in nearby star-forming regions have revealed that nearly all large disks ($\gtrsim$ 50 au in dust continuum emission) host substructures across different
evolutionary stages (e.g., \citealt{2020Andrews}, \citealt{2018Long}). 
These substructures are thought to be related to gas pressure maxima (e.g., \citealt{2012Pinilla}) and they are the locations where dust grains are trapped (e.g., \citealt{2016vanderMarel}), allowing for the formation of planetesimals (e.g.,  \citealt{2020Pinilla}).
Since most of the planets are thought to form in the inner tens of au close to the star, studying the innermost part of the disk is crucial to the understanding of whether and how dust grains flow through the gaps and grow in those
regions. Moreover, the inner regions of disks are host to the processes that influence the global evolution of the disk, such as magnetohydrodynamic (MHD) winds and photoevaporation. In particular, at each evolutionary stage, radially extended MHD winds, surrounding collimated fast jets, {are thought to be the primarily drivers of} disk accretion in the planet formation region ($\sim 1-30$ au), transporting mass inwards and angular momentum outwards ( \citealp{2023PascucciPPVII}, \citealp{2021Lesur}). Evolving towards Class II disks, as the accretion rate falls, the jet disappears, while {models predict that} disk winds persist, evolving from MHD-dominated to phoevaporation dominated and clearing the disk (\citealp{2017ErcolanoPascucci}, \citealp{2023PascucciPPVII}).
This scenario is supported by observations of optical forbidden line of [O I]6300 $\mathring{\mathrm{A}}$, which is a long-established
tracer of outﬂows in T Tauri stars. The high velocity components HVC of [O I]6300 $\mathring{\mathrm{A}}$ clearly originate in high-velocity (micro)jets (e.g., observations in \citealp{2000Dougados}), while the low-velocity components LVC are proposed to possibly track disk winds (\citealp{2014Natta}).
{In line with the weaker emission expected over time, a survey conducted in the Upper Sco region by \citealp{2023Fang} shows a lower detection rate of  HVCs of [O I] emission, while the proportion of single component LVCs is increasing.}
Moreover, the [O I] luminosity is correlated with the luminosity of infrared emission from the disk and with the accretion luminosity, demonstrating that the forbidden line emission is powered by accretion (e.g., \citealp{1990Cabrit}, \citealp{2016ASimon}, \citealp{2019Banzatti}). Thus, this shows that there is a link between the outflow and the stellar accretion. 
Finally, the positive correlation between the centroid shift of the HVC and the accretion luminosity suggests that a stronger accretion drives faster and stronger jets (\citealp{2019Banzatti}).

Transition disks with large inner dust cavities > 20 au are the ideal targets for studying the inner tens of au of protoplanetary disks, as the central emission can be fully disentangled from the outer disk even at moderate angular resolution of $0''.1$ thanks to the high spatial resolution and sensitivity reached with ALMA. The cavities of these disks are not completely devoid of dust and gas and an inner disk often exists close to the star, as shown by the presence of near-infrared excess over the stellar black body (e.g., \citealt{2007Espaillat}) and the high observed accretion rates onto the central stars (e.g., \citealt{2014Manara}).
Moreover, with the high spatial resolution and sensitivity reached with ALMA, CO emission has been detected inside the cavities (e.g., \citealp{2016vanderMarel}) and dusty inner disks with outer radii up to 10 au have been detected in many transition disks in millimeter continuum emission (e.g., \citealt{2020FrancisAndVDMarel}, \citealp{2021Cieza}). However, most of these inner disks are only marginally resolved or unresolved, thus their properties and compositions are still unclear. 

An efficient tool for studying the dust properties in these regions is the analysis of the slope of the spectral energy distribution (SED) at millimeter wavelengths.
If the dust emission is optically thin, values of the spectral index $\alpha$ of the (sub)millimeter emission of $2 <\alpha \leq 3$ indicate dust growth and presence of large dust particles (e.g., \citealt{2001DAlessio}).
Conversely, values between $-0.1 <\alpha < 1.5$ reveal a lack of millimeter dust and indicate that the emission is likely dominated by free-free emission.
Evidence of free-free emission from the innermost region of the disk associated with an ionized radio jet (e.g., \citealt{1986Reynolds}), from a photoionized wind (e.g., \citealt{2012Pascucci}), or both (e.g., \citealt{2016Macias}) has already been found in spatially resolved observations of a handful of single-source studies on TW Hydrae and GM Aur (\citealt{2021Macias}, \citealt{2016Macias}). Moreover, unresolved SED photometry at centimeter wavelengths in other protoplanetary disks have shows evidence of free-free emission (e.g., \citealp{2012Ubach}, \citealp{2017Zapata}), which is shown to be highly variable, although the underlying physical mechanism causing this variability remains unclear (e.g., \citealp{2017Ubach}).

Therefore, a multiwavelength study of the central emission close to the star in a wider sample of transition disks is needed to put more robust constraints on the nature of this emission. In addition, such a study would investigate possible correlations between the emission and the properties of the disk and the star.

In this work, we present a multiwavelength study of 11 transition disks, using archival ALMA observations at 1 mm (Band 7 and 6) and 3 mm (Band 4 and 3), complemented with centimeter (cm) observations reported in the literature conducted with the Karl G. Jansky Very Large Array (JVLA), the Australia Telescope Compact Array (ATCA), and the Green Bank Telescope (GBT).
The paper is structured as follows. In Section \ref{sample}, we present the sample we analyzed in this work. In Sections \ref{analysis}, \ref{results}, and \ref{discussion}, we present our analysis,  results, and a discussion.  In Section \ref{conclusions}, we give our conclusions.

\section{Sample and data reduction}\label{sample}
    \subsection{Target selection}

\begin{table*}
\begin{center}
\tiny
\caption{Sample of objects used this work.}\label{tab:sample}
\begin{tabular}{cccccccccccc}
\hline\hline 
 Target & R.A.  & Decl.  & d & SpT & M$_*$  & log L$_*$ &  log $\text{L}_\text{acc}$&  log $\dot{\text{M}}$  & log $\text{L}_\text{X}$ &inc & Ref. \\
 & (ICRS J2000) & (ICRS J2000) & [pc] & & [M$_\odot$] & [L$_\odot$] & [L$_\odot$] & [M$_\odot$ yr$^{-1}$] & [erg s$^{-1}$] &[deg] & \\
\hline 
 AB Aur      & 04:55:45.8459 & +30:33:04.2921 & 163     & A0   & 2.2   &  $1.61^{+0.19}_{-0.21}$ & $1.32^{+0.14}_{-0.15}$  & $-6.13^{+0.25}_{-0.27}$ & $29.88\pm0.30^\ddag$ & 23 &  1,2,2,5  \\
 GGTau AA/Ab & 04:32:30.3507 & +17:31:40.4941      & 149     &  K7  &  0.7  & $0.20\pm0.20^{*}$ & $0.024\pm0.25$   & $-7.3\pm0.5$ & $29.64\pm0.30^\ddag$ & 36 & 3,..$\dagger$,3,5   \\
 HD100453   &   11:33:05.5766 & -54:19:28.5471    &  104    & A9   & 1.5   &  $0.85^{+0.08}_{-0.09}$ & $-0.35^{+0.28}_{-0.32}$  & $-7.79^{+0.29}_{-0.33}$ & -- & $30$ & 2,2,2  \\
 HD100546     &         11:33:25.4409 & -70:11:41.2413 & 110     & B9   & 2.2   &  $1.45^{+0.11}_{-0.11}$ & $0.97\pm0.14$ & $-6.59^{+0.15}_{-0.12}$ & $29.22\pm 0.30^\ddag$ & 42 & 2,2,2,5  \\
 HD135344B & 15:15:48.4460 &  -37:09:16.0243 & 136 & F5 & 1.6 & $0.94^{+0.13}_{-0.14}$ & $0.09^{+0.21}_{-0.22}$& $-7.25^{+0.21}_{-0.23}$ & -- & 12 & 2,2,2,5 \\
 HD142527 & 15:56:41.8883 & -42:19:23.2483      &  157    & F6   &  2.3  &  $1.39^{+0.09}_{-0.10}$ & $0.65^{+0.18}_{-0.19}$  & $-6.61^{+0.15}_{+0.18}$ & $29.60\pm0.30^\ddag$ & 27 & 2,2,2,5  \\
 HD169142    & 18:24:29.7800 & -29:46:49.3274 & 114     & A5   & 2   & $1.31^{+0.12}_{-0.22}$ & $0.59\pm0.15$  & $-7.09^{+0.26}_{-0.29}$ & $28.90\pm0.30^\ddag$ & $12$ & 1,2,2,5  \\
 MWC758 & 05:30:27.5286 & +25:19:57.0763    & 160     &  A7  &  1.6  &  $1.04^{+0.12}_{-0.08}$ & $0.41^{+0.17}_{-0.18}$  &  $-7.00^{+0.24}_{-0.22}$ & $29.18\pm0.09$ & 21 & 1,2,2,6 \\
 SR24S & 16:26:58.5134 &  -24:45:36.7226   & 114     & K1   & 1.5   &  $0.40\pm0.20^{*}$ & $0.10\pm0.25$ &  $-7.15\pm0.5$ & -- & $46$ & 3,..$\dagger$,3  \\
 TCha  &        11:57:13.5245  & -79:21:31.5305 & 110     & K0   & 1.2   & $0.11\pm0.20^{*}$ & $0.012\pm0.25$    & $-8.4\pm0.5$ & $30.47\pm0.30^\ddag$ & 73 & 3,..$\dagger$,3,5  \\
 WSB60   & 16:28:16.5129 & -24:36:58.0549 & 137     & M6   & 0.2   &  $0.33\pm0.20^{*}$ & $-1.51\pm0.25$  & $-7.93 \pm 0.35$ & -- & 28 & 4,4,4  \\

\hline
\end{tabular}
\end{center}
\tablefoot{Columns show the star coordinates; distance, d;  spectral type, SpT;  stellar mass, M$_*$; and luminosity, L$_*$;  accretion luminosity, $\dot{\text{L;}}$  accretion rate onto the central star, $\dot{\text{M}}$; X-ray luminosity; and the outer disk inclination, inc. The last column reports the references for the luminosity,  accretion luminosity,  accretion rates, and  X-ray luminosity. Distances are from Gaia DR2 (\citealp{2018gaiaCollabDR2}). The inclination of the outer disks are from \citealp{2020FrancisAndVDMarel} and their typical uncertainty is in the range of a few degrees.$^{*}$ The uncertainty on the luminosity for these targets is assumed to be 0.20 dex, based on the typical uncertainty estimated by \cite{2020Wichittanakom}. $\dagger$ The accretion luminosity on these targets is calculated with equation (2) in \cite{2020Wichittanakom}, with uncertainties assumed to be 0.25 dex (\citealp{2023ManaraPPVII}}). $\ddag$ X-ray luminosities derived correcting the fluxes reported in Table 4 in \citealp{2019Dionatos} for the distance of each target and assuming a 0.30 dex uncertainty (\citealp{2007Gudel}).
References: 1) \citealp{2018AVioque}, 2) \citealp{2020Wichittanakom}, 3) \citealp{2020FrancisAndVDMarel}, 4) \citealp{2023ManaraPPVII}, 5) \citealp{2019Dionatos}, 6) \citealp{2023Ryspaeva}.

\end{table*}

The sample analyzed in this paper consists of 11 transition disks within 200 pc and their cavities $>20$ au (Table \ref{tab:sample}; see also \citealp{2023VDMarelTDs}), covering a wide range of stellar parameters. The sample was  selected starting from the transition disk sample as defined in \cite{2023VDMarelTDs}, selecting disks which have already been observed with ALMA  at 1 mm (Band 6 and/or Band 7) and at 3 mm (Band 3 and/or 4) and in which a central point source of continuum emission close to the star was detected previously (e.g., \citealp{2020FrancisAndVDMarel}). The transition disk around DoAr44 is excluded from the sample since, although the detection of a central point source emission is claimed by \cite{2021Cieza}, its flux density is detected at less than $3\sigma$ and is not reported in a more recent study on the same dataset by \cite{2023Arce-Tord}. 

The spectral types of the stars included in the sample vary from B9 to M6, the stellar luminosities range from $0.2$ to $65$ $L_\odot$, the stellar masses range from $0.2$ to $2.3$ $M_\odot$, and accretion rates from $10^{-10}$ to $10^{-6}$ $M_\odot$/yr (e.g., \citealp{2023ManaraPPVII}, \citealp{2020FrancisAndVDMarel}).
Table \ref{tab:sample} reports the stellar properties of all disks in the sample analyzed in this work.
When available, the ALMA data have been complemented with centimeter observations conducted by ATCA and JVLA, with a typical angular resolution from $0''.1$ to $1''-2''$ (\citealp{2014Rodriguez},
\citealp{2014Osorio}, \citealp{2015Casassus}, \citealp{2015Marino}, \citealp{2015Wright}, \citealp{2017Zapata}), or by GBT, with a angular resolution of $24''$ (\citealp{2022GreavesMason}).
The fourth column in Table \ref{tab:flux} reports the Project IDs of the data analyzed in this work and the references for values taken from literature.
In addition to the disks in our sample, we included in the discussion TW Hya and LkCa 15 for which a multiwavelength analysis has already been conducted in literature (\citealp{2021Macias}, \citealp{2020Facchini}).

    \subsection{Data reduction}

\begin{figure*} 
   \centering
   \includegraphics[width=\textwidth]{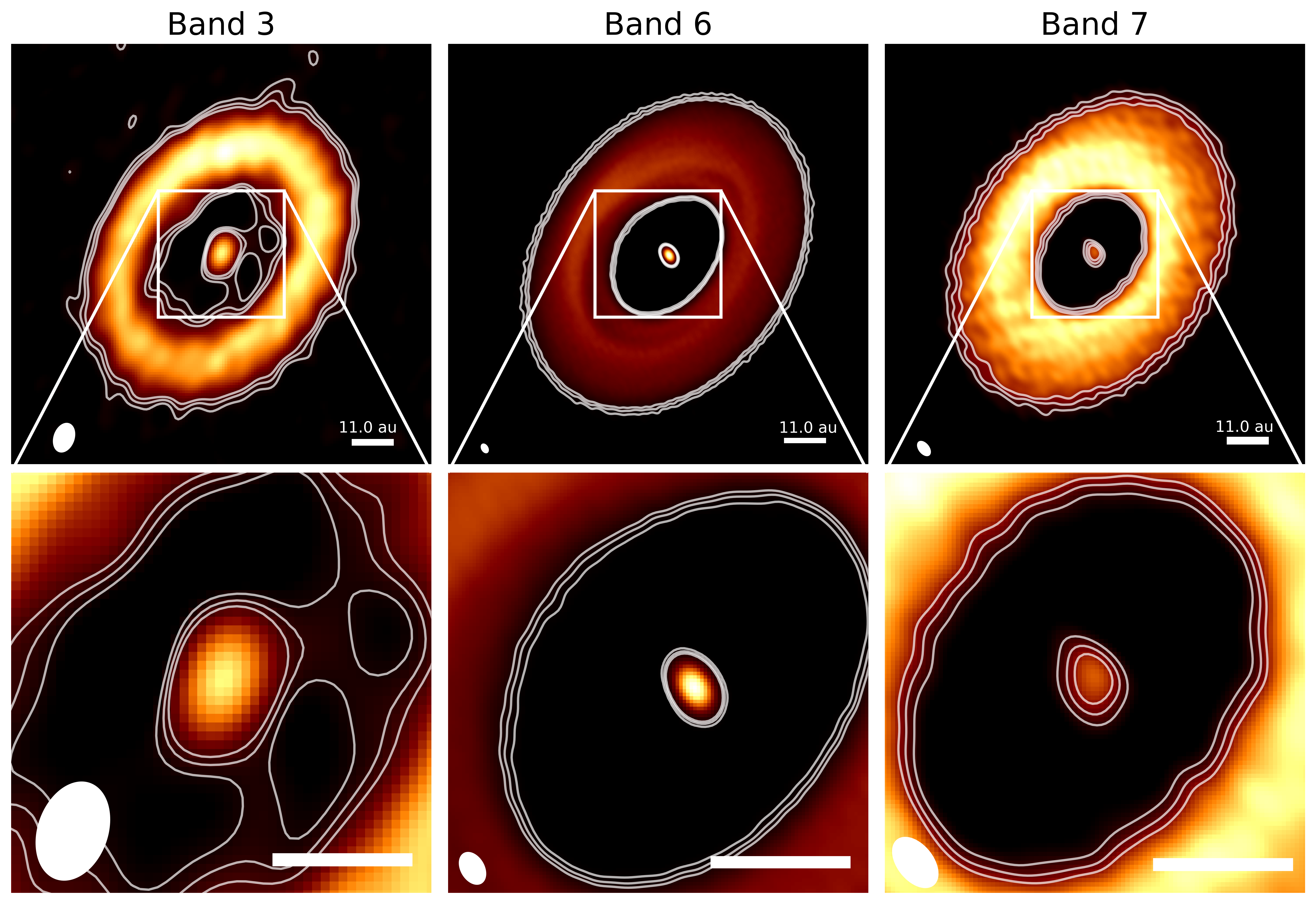}
      \caption{Intensity maps of the disk around HD100546. The first, second, and third columns show the images of the disk in Band 3, Band 6, and Band 7, respectively. The images in the first row are $1''.0 \times 1''.0$, while the zooms in the second row are $0''.3 \times 0''.3$.  In each image, the color scale has the peak flux as the maximum, and the image rms as minimum. All bars in the bottom right are $0''.1$ in length, which is $\sim11$ au at the distance of the source. White contours show three, five, and seven times the rms of the continuum emission.}
    \label{fig:exMap}  
\end{figure*}

The calibrated data were retrieved using the ALMA Pipeline reduction scripts provided by the observatory and no self-calibration process was applied to the data. Continuum images of each target were created with the task \texttt{tclean} of the Common Astronomy Software Applications package (CASA, version 5.7.0, \citealp{2007AMcMullinCASA}). 
As the relative sensitivity and spatial sampling vary widely across the sample, for each individual disk and observing wavelength the images were produced with different weighting scheme. The cleaning was performed using natural, uniform, and Briggs (with robust parameter between 0.5 and -1.0) weighting and selecting (for each image) the weighting that maximizes the signal-to-noise ratio (S/N) for the inner disks (see also \citealp{2020FrancisAndVDMarel}). Table \ref{tab:Clean} reports for each disk and each wavelength the adopted weighting scheme, the final beam size, and rms of the image. Figure \ref{fig:exMap} shows, as an example, the intensity maps of the continuum emission in the disk around HD100546. The intensity maps of all disks can be found in Appendix \ref{app:propImages} (Figures \ref{fig:HD100453}-\ref{fig:ABaur}).

     \begin{table*}
\begin{center}
\small
\caption{Image properties}\label{tab:Clean}
\begin{tabular}{cccccc}
\hline\hline
  Target & Frequency & Weighting & rms & Beam Size & Beam PA\\

  & [GHz] & Scheme &  [mJy/beam] & & [deg]  \\
\hline

 GG Tau AA/Ab & \m{ 97.50 \\ 236.51 \\ 343.51}
     &  \m{ Uniform \\ Briggs, r=0.0  \\ Uniform }
     &  \m{ 0.034 \\ 0.090 \\ 0.16}
     &  \m{ $0''.515 \times 0''.380$  \\ $0''.249 \times 0''.211$  \\ $0''.195 \times 0''.135$ }
     &  \m{ -32.533 \\ 22.565 \\ -32.270}
     \\ [0.6cm]    
     
 HD100546 & \m{98.39 \\ 225.00 \\ 338.18}
    &  \m{Briggs, r=0.0\\ Briggs, r=0.0 \\ Briggs, r=0.0}
    &  \m{ 0.060  \\ 0.0081 \\ 0.096 }
     &  \m{  $0''.073 \times 0''.050$ \\ $0''.025 \times 0''.017$ \\ $0''.042 \times 0''.026$ }
    &  \m{ -19.591  \\ 30.372 \\ 38.642 }
      \\ [0.8cm]

 HD142527 & \m{ 103.36 \\ 144.98 \\ 225.05 \\ 343.47}
    &  \m{ Briggs, r=-1.0 \\ Briggs, r=0.0  \\ Uniform \\ Uniform}
     &  \m{  0.039 \\ 0.030 \\  0.40 \\ 0.19}
     &  \m{  $0''.329 \times 0''.298$ \\ $0''.233 \times 0''.170$ \\ $0''.243 \times 0''.220$ \\ $0''.139 \times 0''.120$ }
     &  \m{  58.830 \\ 89.304 \\  47.997 \\ -51.061}
     \\[0.8cm]

  AB Aur & \m{ 144.76 \\ 232.71 \\ 337.77}
    &  \m{ Natural \\ Briggs, r=0.5 \\ Briggs, r=0.5 }
    &  \m{ 0.0083  \\  0.020 \\ 0.18}
     &  \m{ $0''.117 \times 0''.066$  \\ $0''.040 \times 0''.022$  \\ $0''.310 \times 0''.170$ }
     &  \m{ 21.634  \\  22.676 \\ 30.189}
     \\ [0.8cm] 

HD100453 & \m{96.15 \\ 225.71}
    &  \m{Briggs, r=0.5 \\ Natural }
     &  \m{ 0.017 \\ 0.021 }
     &  \m{$0''.114 \times 0''.086$\\ $0''.035 \times 0''.031$ }
     &  \m{ 8.158 \\ 9.183 }
      \\ [0.5cm]

 WSB60 & \m{102.42 \\ 225.00 \\ 343.51}
    &  \m{ Briggs, r=0.0 \\ Natural \\ Briggs, r=0.0}
     &  \m{ 0.049 \\ 0.028 \\ 0.36}
     &  \m{ $0''.187 \times 0''.087$\\ $0''.048\times 0''.030$ \\ $0''.154 \times 0''.128$ }
      &  \m{ 58.126 \\ 88.640 \\ 59.146}
      \\ [0.8cm]

 HD169142 & \m{97.49 \\ 225.0 \\ 338.18} 
    &  \m{ Briggs, r=1.0 \\ Natural  \\ Natural}
    &  \m{ 0.0045  \\ 0.018 \\ 0.13 }
   &  \m{ $0''.055 \times 0''.051$ \\ $0''.072 \times 0''.043$ \\ $0''.193 \times 0''.148$ }
     &  \m{ 65.408  \\ 87.977 \\ 85.637 }
      \\[0.8cm]    
 HD135344B & \m{109.03\\ 155.0 \\ 338.80}
    &  \m{ Natural \\ Briggs, r=0.5 \\ Uniform }
    &  \m{ 0.024 \\ 0.013 \\ 0.31}
     &  \m{  $0''.110 \times 0''.075$ \\ $0''.077 \times 0''.046$ \\ $0''.245 \times 0''.180$   }
      &  \m{ 28.087 \\ 35.248 \\ 52.532}
      \\[0.5cm]    

 MWC758   &  \m{ 343.5 } &
            \m{Briggs, r=0.5}  &  
            \m{ 0.019}  &  
            \m{$0''.040 \times 0''.031$}  &
            \m{ 0.515}
             \\ [0.3cm]   
              
     SR24S & \m{ 108.84 \\ 225.00 \\ 343.48}
    &  \m{ Briggs, r=-1.0 \\ Natural \\ Superuniform }
     &  \m{ 0.037 \\  0.037  \\ 0.35}
      &  \m{ $0''.078 \times 0''.065$ \\  $0''.043 \times 0''.035$  \\ $0''.133 \times 0''.107$}
      &  \m{ 67.058 \\  85.702  \\ 66.700}
       
      \\[0.1cm]

  TCha & \m{97.49 \\ 225.17 \\ 338.12} 
     &  \m{ Briggs, r=-1.0\\ Natural \\ Superuniform }
     &  \m{ 0.029 \\ 0.082  \\ 0.40}
     &  \m{ $0''.073 \times 0''.033$\\ $0''.155 \times 0''.086$  \\ $0''.296 \times 0''.170$ }
      &  \m{ -20.278 \\ -11.487  \\ 31.430}
    \\     
    \hline

\hline 
\end{tabular}
\end{center}
\tablefoot{ Columns show for each target the observing frequency, the selected weighting scheme for the cleaning, the rms of the final image, the beam size, and the beam position angle (PA).
}
\end{table*}

\section{Analysis}\label{analysis}

\subsection{Central emission modeling and outer disk flux density}

For all disks in the sample, central emission close to the star has been detected with ALMA both at 3 mm (Band 3 and/or Band 4) and at 1 mm (Band 6 and/or Band 7), with the exception of T Cha, which was only detected in Band 3 and non-detected at 1 mm. The central emission is unresolved or marginally resolved in all cases. Previous claims of spatially resolved inner disk emission by \cite{2020FrancisAndVDMarel} are potentially caused by the influence of sidelobes inside the cavity, resulting from the emission of the outer disk, on the outcome of the CASA \texttt{imfit} task, for the marginally resolved cases. The only exception is the central emission in WSB60, which deconvolved size is $160-250 \times 140-200$ mas -- depending on the observing wavelength, similar to what was previously recovered (\citealp{2020FrancisAndVDMarel}). 
  
For each disk and each wavelength, the flux density of the detected central emission $F_\text{inner}$ is estimated through a 2D Gaussian fitting with the CASA \texttt{imfit} task. If no central emission close to the star is detected down to three times the rms level of the image and the cavity is resolved, an upper limit of 5 times the rms is adopted.
The total flux density $F_\text{tot}$ of the disk is measured with aperture photometry, summing all the pixels with emission higher than three times the rms level in a circular region with a radius larger than the disk. 
The outer disk flux density $F_\text{outer}$ is then estimated subtracting $F_\text{inner}$ from $F_\text{tot}$.
Table \ref{tab:flux} reports the estimated flux densities $F_\text{inner}$ and $F_\text{outer}$. The uncertainties on the flux densities take into account the 10\% and 5\% calibration uncertainty usually assumed on ALMA flux measurements in Bands 6/7 and Bands 3/4, respectively. 
If no inner disk is detected, the total flux density of the disk is assumed as an estimate of the outer disk flux density. 

\begin{table*}
\begin{center}
\small
\caption{Observed flux densities in each band}\label{tab:flux}
\begin{tabular}{clllll}
\hline\hline
 Target & Frequency& References/ & Total Flux & Inner Flux & Outer Flux \\

 & [GHz] & Project IDs and PI &  [mJy] & [mJy] & [mJy] \\
\hline

 GG Tau AA/Ab & \m{31.0 \\ 97.50 \\ 236.51 \\ 343.51}
     &  \m{1, GBT \\ 2018.1.00618.S, PI:Tang \\ 2018.1.00532.S, PI:Denis Alpizar \\ 2018.1.00618.S, PI:Tang}
     &  \m{$1.05\pm0.04$ \\ $58 \pm 6$ \\ $621 \pm 62$ \\ $1337 \pm 134$}
     &  \m{- \\ $2.1 \pm 0.2$ \\ $9.3 \pm 1.0$ \\ $21.0 \pm 2.2$}
     &  \m{-\\$56.0 \pm 5.6$ \\ $612 \pm 61$ \\ $1316 \pm 132$}
     \\ [0.8cm]    
     
 HD100546 & \m{4.8\\8.64 \\98.39 \\ 225.00 \\ 338.18}
    &  \m{2, ATCA \\ 2, ATCA \\ 2017.1.00885.S, PI:Kama \\ 2018.1.01309.S, PI:Perez \\ 2015.1.00806.S, PI:Pineda}
    &  \m{$0.38\pm0.12^{\ddag}$\\$0.83\pm0.44^{\ddag}$\\ $75 \pm 8$ \\ $438 \pm 44$ \\ $1211 \pm 121$}
     &  \m{- \\- \\ $1.8 \pm 0.3$ \\ $3.4 \pm 0.5$ \\ $5.6 \pm 0.8$}
     &  \m{-\\-\\$73 \pm 8$ \\ $434 \pm 43$ \\ $1205 \pm 121$}
      \\ [1cm]

 HD142527 & \m{34.0 \\ 103.36 \\ 144.98 \\ 225.05 \\ 343.47}
    &  \m{3, ATCA \\ 2013.1.00670.S, PI:Momose \\ 2017.1.00987.S, PI:Kataoka \\ 2015.1.01353.S, PI:Christiaens \\ 2015.1.00425.S, PI:Kataoka}
     &  \m{- \\ $57 \pm 6$ \\ $236 \pm 24$ \\ $858 \pm 86$ \\ $2960 \pm 296 $}
     &  \m{$0.082 \pm 0.022$ \\ $0.49 \pm 0.07$ \\ $0.81 \pm 0.15$ \\ $1.6 \pm 0.5$ \\ $3.8 \pm 0.5 $}
     &  \m{ - \\$57 \pm 6$ \\ $235 \pm 24$ \\ $856 \pm 86$ \\ $2956 \pm 296$}
     \\[1cm]

  AB Aur & \m{8.9  \\ 144.76 \\ 232.71 \\ 337.77}
    &  \m{4, VLA \\ 2019.1.00579.S, PI:Fuente \\ 2015.1.00889.S, PI:Tang \\ 2012.1.00303.S, PI:Tang}
    &  \m{ - \\ $23.2 \pm 2.3$ \\ $9.5 \pm 2.6^{\dagger}$ \\ $158 \pm 16$}
     &  \m{ $0.136 \pm 0.06$\\ $1.27 \pm 0.13$ \\ $1.95 \pm 0.22$ \\ $1.31 \pm 0.29$}
     &  \m{ - \\ $21.9 \pm 2.2$ \\ $7.6 \pm 2.6$ $^\dagger$ \\ $157 \pm 16$}
     \\ [1cm]

 HD100453 & \m{96.15 \\ 225.71 \\ 338.80}
    &  \m{2017.1.01678.S, PI:van der Plas \\ 2017.1.01678.S, PI:van der Plas \\ 5, ALMA}
    &  \m{$19.0 \pm 2.0$ \\ $185 \pm 19$ \\ $510 \pm 50$}
     &  \m{$0.26 \pm 0.03$ \\ $0.81 \pm 0.10$ \\ $1.60 \pm 0.16$}
     &  \m{$18.7 \pm 1.9$ \\ $184 \pm 19$ \\ $508.4 \pm 50.8$}
      \\ [1.cm]

 WSB60 & \m{9.0 \\ 102.42 \\ 225.00 \\ 343.51}
    &  \m{6, VLA  \\ 2016.1.01042.S, PI:Chandler \\ 2018.1.00028.S, PI:Cieza \\ 2013.1.00157.S, PI:Looney}
    &  \m{ $<0.026$ \\$10.7 \pm 1.2$ \\ $85 \pm 9$ \\ $189 \pm 19$}
     &  \m{ - \\$4.9 \pm 0.5$ \\ $23.3 \pm 2.6$ \\ $41 \pm 5$}
     &  \m{ - \\$5.9 \pm 0.9$ \\ $62 \pm 7$ \\ $148 \pm 16$}
      \\ [1cm]

 HD169142 & \m{5.5 \\ 9.0 \\ 9.5 \\ 33.0 \\ 44.0 \\97.49 \\ 225.0 \\ 338.18} 
    &  \m{7, VLA \\ 7, VLA \\ 8, VLA \\ 8, VLA \\ 8, VLA \\ 2018.1.01716.S, PI:Mac{\'\i}as \\  2016.1.00344.S, PI:Perez \\ 2012.1.00799.S, PI:Honda}
    &  \m{  - \\ - \\ $0.050\pm0.010$  \\ $0.850\pm0.150$ \\ $2.0\pm0.4$ \\ $16.1\pm1.6$ \\ $154\pm15$ \\ $536\pm54$}
     &  \m{$<0.036$ \\ $<0.039$ \\ $0.020\pm0.005$\\ $0.045\pm0.014$\\ $0.074\pm0.015$ \\$0.090 \pm 0.011$ \\ $0.25 \pm 0.05$ \\ $<0.63$}
     &  \m{- \\ - \\ - \\ - \\ - \\ $16.1 \pm 1.6$ \\ $154 \pm 15$ \\ $536 \pm 54$}
      \\[1.5cm]    

 HD135344B & \m{ 9 \\ 109.03 \\ 155.0 \\ 225.0 \\ 338.80}
    &  \m{ 6, VLA \\ 2016.1.00340.S, PI:Cazzoletti  \\ 9, ALMA \\ 2012.1.00158.S, PI:van Dishoeck }
    &  \m{ $<0.100$ \\ $10.6 \pm 0.9$ \\ $38.7 \pm 3.9$ \\ $105.1\pm0.3$ \\ $628 \pm 63$}
     &  \m{ - \\ $<0.12$ \\$0.0780 \pm 0.0096$ \\ $0.120\pm0.024$ \\ $<1.56$ }
     &  \m{ - \\ $10.6\pm 0.9$\\$38.58 \pm3.9 $\\ $104.98 \pm 0.30$ \\ $628\pm63$ }
      \\[0.8cm]    

 MWC758   &  \m{33.0 \\ 343.5 } &
            \m{10, VLA \\ 2017.1.00492.S, PI:Dong}  &  
            \m{ $0.366\pm0.049$ \\ $185\pm18$}  &  
            \m{$0.077\pm0.059$ \\ $0.23\pm0.03$}  &    
            \m{$0.289\pm0.33$ \\ $184\pm18$}
             \\ [0.5cm]   
              
     SR24S & \m{ 8.9 \\ 108.84 \\ 225.00 \\ 343.48}
    &  \m{6, VLA  \\ 2017.1.00884.S, PI:Pinilla \\ 2018.1.00028.S, PI:Cieza \\ 2013.1.00157.S, PI:Looney}
    &  \m{ $0.096 \pm 0.07$\\ $29 \pm 3$ \\ $179 \pm 18$ \\ $473 \pm 48$}
     &  \m{ - \\ $0.42 \pm 0.05$ \\ $0.71 \pm 0.08$ \\ $<1.74$}
     &  \m{ - \\ $29 \pm 3$ \\ $178 \pm 18$ \\ $473 \pm 48$}
      \\ 
      [1cm]

 TCha & \m{5.5 \\ 97.49 \\ 225.17 \\ 338.12} 
     &  \m{11, ATCA \\ 2015.1.00979.S, PI:Pascucci \\ 2017.1.01419.S, PI:Caceres \\ 2012.1.00182.S, PI:Brown}
     &  \m{ - \\ $12.3\pm1.4$ \\ $93\pm9$ \\ $213\pm22$}
     &  \m{ $0.3\pm0.1$\\ $0.45 \pm 0.04$ \\ $<0.41$ \\ $<2.0$}
     &  \m{- \\ $11.8 \pm 1.4$ \\ $93 \pm 9$ \\ $212 \pm 22$}
    \\[0.5cm]

\hline 
\end{tabular}
\end{center}
\tablefoot{$^\dagger$The short baseline configuration for Band 6 (232.71 GHz) observations of AB Aur is missing. For this reason, the estimated total flux is a lower limit, and thus it is not included in the power-law fitting. $^{\ddag}$ The emission at 3.5 cm and 6.2 cm is shown to be variable by \cite{2015Wright}. The flux reported here and used in the analysis is the average of the measured fluxes in different epochs.
As for the cm-wave data taken from literature, flux densities values are reported under the Inner Flux column if the outer disk emission is disentangled from the inner disk emission, while on the Total Flux column if the two emissions are not disentangled.
References for literature data: 1) \citealp{2022GreavesMason}, 
2) \citealp{2015Wright}, 3) \citealp{2015Casassus}, 4) \citealp{2014Rodriguez}, 5) \citealp{2020Rosotti}, 6) \cite{2017Zapata}, 7) \citealp{2014Osorio}, 8) \citealp{2017Macias}, 9) \citealp{2021Casassus}, 10) \citealp{2015Marino}, 11) \citealp{2012Ubach}.
 
 }

\end{table*}

\subsection{Millimeter spectral index}

\begin{figure*} 
   \centering
   \includegraphics[ width=\textwidth]{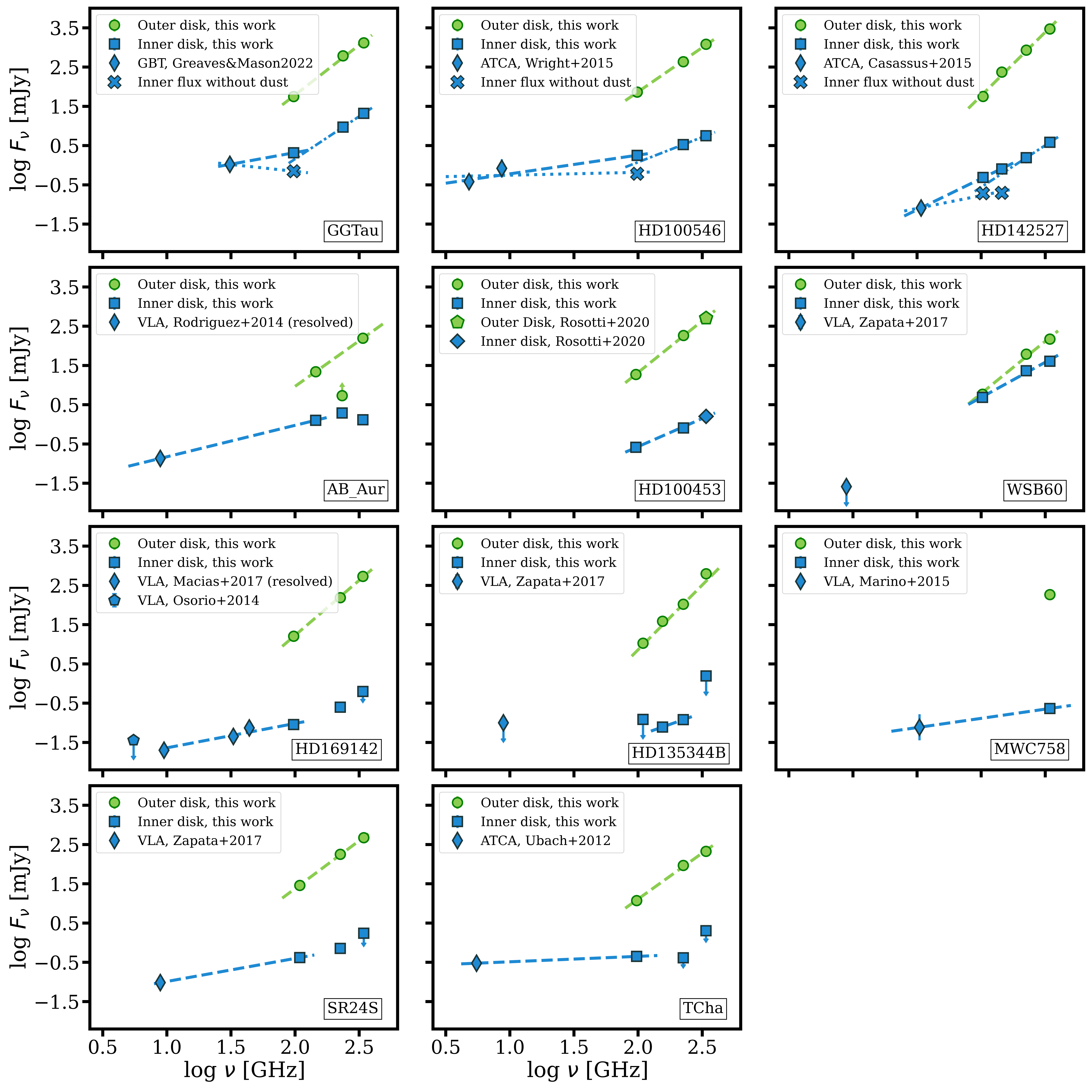}
      \caption{SED of each disk in the sample. Green dots show the estimated fluxes at different wavelengths for the outer disk, while blue squares refer to the inner disk. 
      {As for the  cm data from the literature, if the emission of the inner disk is resolved from the emission from the outer disk, the label ``resolved'' is added to the legend.}
      The green lines show the best-fit for the outer disk. The blue lines refer to the inner disk emission. In particular, dash-dotted, dashed, and dotted blue lines in the first row refer to the best-fit between the 1 mm observations, the best-fit between the 3mm and cm observations, and the best-fit between the 3 mm data without the dust contribution and the cm-data, respectively. The dashed blue lines in the second, third, and fourth rows show the best-fit between all the available data. See the text for further details.
      }
    \label{fig:SED}  
\end{figure*}

Figure \ref{fig:SED} shows the SED for each disk in the sample.
The spectral index of the emission has been estimated for both the central emission and the outer emission, with a linear fit, given an assumed correlation between the flux density and frequency:
\begin{equation}\label{eq:corrF}
    \log F_\nu = \log A + \alpha \log \nu.
\end{equation}
Since the outer disk emission is detected and resolved in all the disks at each analyzed wavelength, the spectral index of the outer disk, $\alpha_{\text{outer}}$, is estimated including all the available data in the linear fit. If more than two detections are available, the best-fit parameters -- $\log A_\text{outer}$ and $\alpha_\text{outer}$ -- are estimated using the \texttt{curve\_fit} function of the \texttt{scipy} module; otherwise, the spectral index (and the relative intercept) is ``manually'' calculated as $\alpha_\text{outer} = \log(F_{1}/F_{2})/ \log (   \nu_{1}/\nu_{2})$, where $F_1$ and $F_2$ are the observed flux densities at the frequency $\nu_1$ and $\nu_2$, respectively. The best-fit relation is shown in Figure \ref{fig:SED} with green dashed lines.
The spectral index of the central emission $\alpha_{ \text{inner}}$ is estimated with a slightly different analysis based on the bands in which the central emission is detected.

\begin{figure} 
   \centering
   \includegraphics[width=0.5\textwidth]{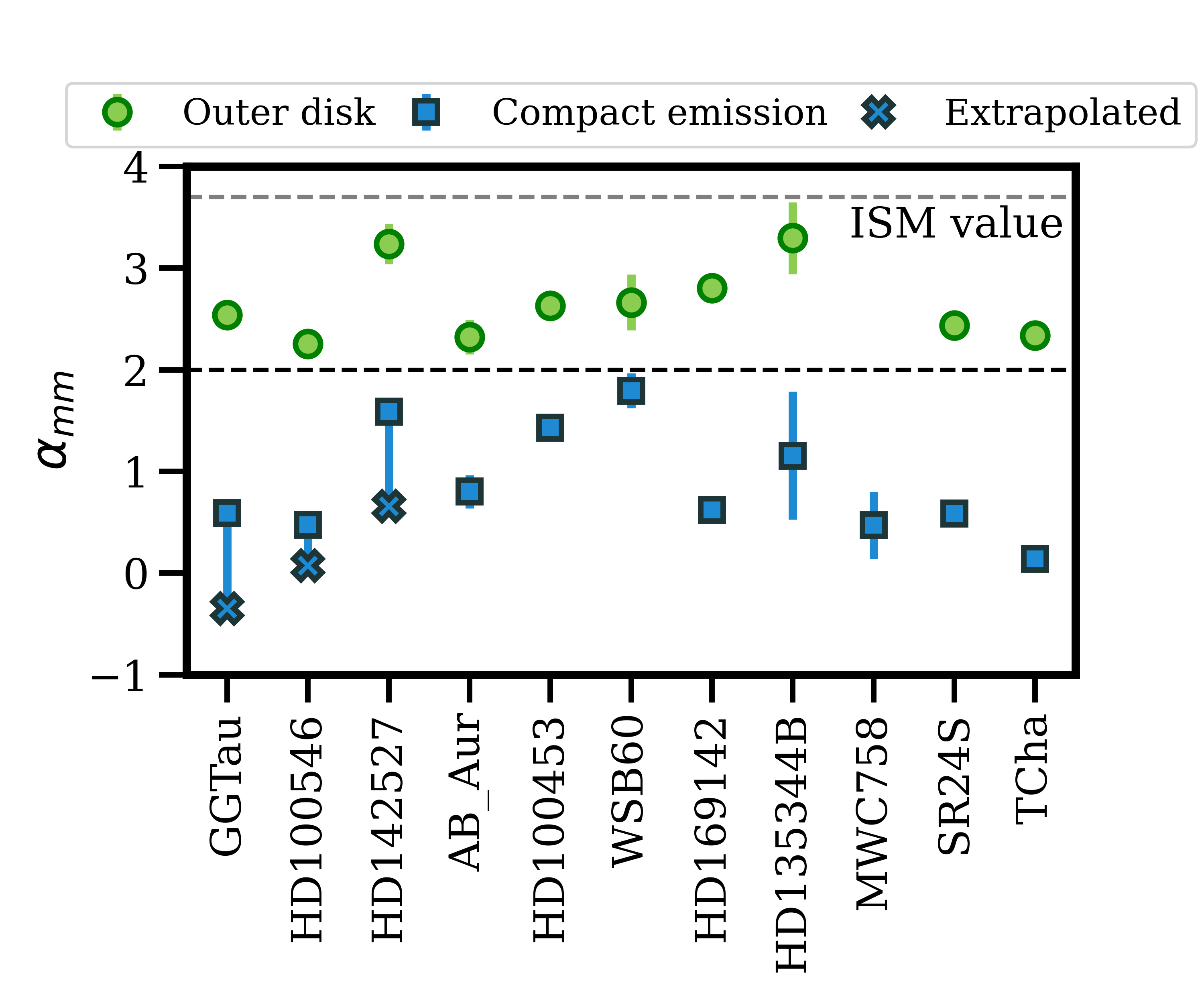}
      \caption{Spectral indices of the mm-wave emission. Green dots refer to the outer disk, while blue squares and red crosses refer to the central  emission. For the first three targets -- GGTau, HD100546, and HD142527 -- the red crosses indicate $\alpha_\text{extrap}$, while the blue squares indicate $\alpha_\text{3mm-cm}$ (see text for further details). The dashed grey line shows the value of the spectral index in the ISM, $\alpha=3.7$, while the black line shows $\alpha=2$.}
    \label{fig:spectralIndex}  
\end{figure}

As for GG Tau, HD100546, and HD142527, for which the central emission is detected in both Bands 6 and 7, we first estimated the expected contribution of the thermal dust at 3 mm. Assuming that the emission at 1 mm is dominated by thermal dust with only a small contribution from the free-free emission, we calculated the spectral index of the emission at 1 mm $\alpha_{1mm} = \log(S_{B6}/S_{B7})/ \log (   \nu_{B6}/\nu_{B7})$ (see dashed-dotted lines in the figure) and then we extrapolated the expected dust contribution to the flux density at 3 mm, using Equation (\ref{eq:corrF}). Subtracting the dust contribution thus calculated from the observed flux density at 3 mm, we estimate the fraction of the observed flux density at 3 mm likely associated with free-free emission (blue crosses in the figure). We then estimated the spectral index $\alpha_\text{extrap}$ including the extrapolated flux density at 3 mm and literature data at centimeter wavelengths (see the dotted line in the figure);
$\alpha_\text{extrap}$ is the expected spectral index of the central emission if we assume that the emission at 1 mm is dominated by thermal dust emission. However, an unknown fraction of the 1 mm flux is due to free-free emission, and, thus, $\alpha_\text{extrap}$ is the expected minimum value for the spectral index. To estimate the maximum value of the spectral index, we calculated the spectral index of the 3 mm and cm emission $\alpha_\text{3mm-cm}$ without subtracting the expected contribution of the dust at 3 mm (see blue dashed lines in the figure). Then, $\alpha_\text{3mm-cm}$ is still contaminated with the dust contribution at 2-3 mm; thus, it is the maximum possible value for the spectral index of the central emission. In this way, the interval $[\alpha_\text{extrap},\alpha_\text{3mm-cm}]$ is the best estimate for the spectral index of the inner central emission.
In the case of AB Aur, since $\alpha_{1mm}$ is negative due to missing short baselines and, thus, we cannot estimate the dust contribution in Band 4, we assumed $\alpha_\text{3mm-cm}$ as the best estimate of $\alpha_\text{inner}$.
As for HD100453 and WSB60, although the central emission is detected in both Bands 6 and 7,  $\alpha_\text{inner}$ is estimated using all available data without extrapolating the dust contribution at 3 mm, since no cm-wave data are available to apply the same analysis done for GG Tau, HD100546, and HD142527.
For all the other disks, since the central emission is detected in only one band at 1 mm and, thus, we cannot estimate the dust contribution at lower frequencies,  $\alpha_\text{inner}$, is estimated with all available data. In the case of SR24S {and HD169142}, since we have both Band {3/4} and cm-wave detections, we did not include Band 6 observations in the fit, so that the dust contribution to the estimated spectral index could be minimized.
The estimated spectral indices are summarized in Table \ref{tab:alpha} and shown in  Figure \ref{fig:spectralIndex}.

\begin{table*}
\begin{center}
\small
\caption{Estimated spectral indices for the outer and inner central emission.}\label{tab:alpha}
\begin{tabular}{ccccccc}
\hline\hline 
 Target &  $\alpha_\text{1mm}$ &  $\alpha_\text{3mm-cm}$ &  $\alpha_\text{extrap}$ &  $\alpha_\text{inner}$ &  $\alpha_\text{outer}$ & $\dot M_\text{i} [M_ \odot /\text{yr}]$\\ \hline

 GG Tau & $2.17\pm0.40$ & $0.59 \pm 0.09 $  & $\sim -0.35$ & $[-0.35,0.69](0.17)^{*}$ & $2.54\pm 0.15$ & $(2.45 \pm 3.03)10^{-9}$\\
 HD100546 & $1.27\pm0.49$ & $0.48 \pm 0.09$ & $\sim0.073$& $[0.07,0.57](0.32)^{*}$& $2.25 \pm 0.09$ & $(3.15 \pm 1.20)10^{-9}$\\
 HD142527 & $2.13\pm0.86$ & $1.59 \pm 0.04$ & $\sim0.66$& $[0.66,1.63](1.14)^{*}$& $3.24\pm0.20$ & $(0.675 \pm 1.23)10^{-8}$\\
 AB Aur & -- & $0.80\pm0.16$& -- & $<0.80\pm0.16$& $2.32\pm0.17$ & $(65.94 \pm 9.50)10^{-10}$\\
 HD100453 & -- & --  & -- & $<1.43\pm0.08$ & $2.63\pm0.04$ & --\\
 WSB60 & -- & --  & -- & $<1.79\pm0.17$&  $2.66\pm0.27$ & -- \\

 HD169142 & -- & --  & -- & $<0.62\pm0.11$ & $2.80\pm0.09$ & $(6.30\pm 0.87)10^{-10}$\\
 HD135344B & -- & --  & -- & $<0.96\pm0.71$ & $2.69\pm0.27$ & $(1.09 \pm 2.66)10^{-9}$\\
 MWC758 & -- & --  & -- & $<0.47\pm0.33$ &  -- & $(91.19 \pm 8.64)10^{-11}$ \\ 
 SR24S & -- & --  & -- & $<0.59\pm0.05$ &  $2.45\pm0.04$& $ (11.67 \pm 1.13)10^{-10}$ \\
TCha & -- & --  & -- & $<0.14\pm0.12$ & $2.40\pm0.06$ & $(5.30\pm1.7)10^{-10}$ \\

\hline

\end{tabular}
\end{center}
\tablefoot{Columns show the target name; 1mm spectral index, $\alpha_\text{1mm}$;  spectral index between the 3mm and cm data, $\alpha_\text{3mm-cm}$;  extrapolated spectral index between the 3mm flux without the dust contribution and the cm data, $\alpha_\text{extrap}$;  spectral {index} of the central emission, $\alpha_\text{inner}$; and the spectral index of the outer disk emission, $\alpha_\text{outer}$.
$^{*}$ {We report the confidence interval $[\alpha_\text{extrap},\alpha_\text{3mm-cm}]$ for the spectral index of the central emission, while the average value in the interval that we used to calculate the ionized mass loss rate is reported in brackets}. See the text for further details.}

\end{table*}

\section{Results}\label{results}

As shown in Figure \ref{fig:spectralIndex}, the estimated spectral indices for the outer disk emission are (in all cases) equal to $>2$, with a median value of $\sim2.6$.
The spectral indices of the outer disks are thus consistent with thermal dust emission associated with dust growth, similar to other protoplanetary disks (\citealp{2014Testi}). 
Conversely, values of $\alpha <2$ are found for the central emission close to the star, suggesting that free-free emission may dominate in this region of the disk.
Dust self-scattering effects are known to lower the spectral index down to $\alpha \sim 1.5$ (\citealp{2020Sierra}). However, once we take into account the dust contribution, we expect the mean values of $\alpha_\text{inner}$ to be lower than those reported here, strengthening the indication of free-free emission.  
It is only in two cases, namely, the disk around WSB60 and HD135344B,  that the estimated spectral indices, $\alpha_\text{inner}$, are consistent with $\alpha=2$ within $2\sigma$. As for the disk around WSB60, it is the only case in the analyzed sample where the detected central emission is resolved with a size of $\sim 0''.2$ ($\sim 27.4$ au) and a beam size of $\sim 0''.1$;   thus, it is expected that the emission is most likely dominated by thermal dust emission with $\alpha\sim2$. As for HD135344B, the central emission is not resolved at a resolution of $\sim 0''.1$ ($\sim 13.6$ au) and it is detected only in Bands 4 and 6 with solely an upper limit at cm wavelengths available from the literature (\citealp{2017Zapata}). The estimated spectral index is thus contaminated by the dust contribution and, for this reason, we expect the `real' value of the spectral index to be smaller than the one calculated here, and thus more consistent with free-free emission than with thermal dust.

\section{Discussion}\label{discussion}
While the spectral indices of the outer disk emission are consistent in all cases with thermal dust emission, in 10 out of the 11 disks, the inner central emission appears dominated by free-free emission. This emission can be associated with gas from an ionized jet (e.g., \citealp{1986Reynolds}), a disk wind (e.g., \citealp{2012Pascucci}), or from both (e.g., \citealp{2016Macias}).

\subsection{Inner-disk dust masses}

As we note earlier in this work, the relative lack of mm-dust grains in the majority of inner disks suggests that either such dust grains have drifted quickly towards the central star, that grain growth is less efficient in the inner disk, or that grains grow rapidly to planetesimal sizes in the inner disk. This result calls into question some of the results discussed in \cite{2020FrancisAndVDMarel}, as the inner disk mm-dust mass may be much lower than previously derived at least in the disks analyzed here. In the assumption of optically thin emission and in the blackbody approximation -- namely, ignoring scattering effects, the dust mass is (see the approach described in Section 3.2 and Equation (12) in \citealp{2020FrancisAndVDMarel}):

\begin{equation}
    M_\text{dust} = \frac{F_\nu d^2 \cos i}{\kappa_\nu \frac{2\nu^2k_\text{B}}{c^2} \sqrt[4]{\frac{\phi L_*}{8 \pi \sigma_\text{B}}}} \frac{(R_\text{id}-R_\text{subl})}{2(R_\text{id}^{1/2}-R_\text{subl}^{1/2})},
\end{equation}
where the dust opacity $\kappa_\nu$ is 10 cm$^2$ g$^{-1}$ at 1000 GHz, with an opacity power-law index of $\beta=1.0$ (\citealp{1991BeckwithANDSargent}), $\sigma_\text{B}$ as the Stefan-Boltzmann constant, and $k_\text{B}$ as the  Boltzmann constant. Also, $F_\nu$ is the flux observed at the frequency $\nu$, $i$ is the disk inclination, $d$ the distance, $\phi$ is the flaring angle (taken as 0.02), and $L_*$ the stellar luminosity. The sublimation radius is $R_\text{subl} = 0.07  \sqrt{L_*(L_\odot)}$ (\citealp{2001Dullemond}), while $R_\text{id}$ is the inner disk radius. 
Since the free-free emission is likely to be negligible and not-extended at 1 mm, the inner dust masses are estimated starting from the fluxes observed at Bands 6 or 7 (when available), assuming as an upper limit on $R_\text{id}$ the corresponding beam size. It is only in the case of TCha that (since the central emission is not detected in Bands 6 or 7) we used the flux observed in Band 3 to estimate the mass.
Table \ref{tab:masses} reports for each target the band which flux has been used for the mass estimate, the assumed inner disk radius and the derived inner dust mass.
The millimeter inner disk dust masses estimated {in this way} range between $0.004 M_\oplus$ and $1.37 M_\oplus$.

{These dust masses are highly uncertain. On one hand, within the assumption that the emission is optically thin, the values should be considered as upper limits for all the sources (excluding GG Tau, HD100546, and HD142527) since we cannot distinguish the thermal dust contribution from the free-free emission. On the other hand, if the emission is optically thick, these values should be considered as lower limits. Moreover, the $\alpha_\text{1mm}<3$ for  GG Tau, HD100546, and HD142527 (see Table \ref{tab:alpha}) suggest that scattering effects cannot be neglected.} A full radiative transfer modeling approach is thus required to obtain a better estimate of the inner disk dust masses. This modeling is beyond the scope of this study and left for future work.

\begin{table*}
\begin{center}
\caption{Estimated inner disk dust mass}\label{tab:masses}
\begin{tabular}{ccccc}
\hline\hline 
Target &  Band & Beam size & $R_\text{id}$ [au] &  $M_\text{dust}$ [$M_\oplus$] \\ \hline

GG Tau & Band 7  & $0''.195 \times 0''.135$ & $<30$ & $\sim1.00$ \\
HD100546 & Band 6 & $0''.025 \times 0''.017$ & $<1.9$ &$ \sim0.024$\\
HD142527 & Band 7 & $0''.139 \times 0''.120$ & $<18.8$ & $\sim0.096$ \\
AB Aur & Band 6 & $0''.040 \times 0''.022$  & $<6.5$ & $\sim0.022$ \\
HD100453 & Band 7 & $0''.035 \times 0''.031$ & $<3.6$ & $\sim0.011$  \\
WSB60 & Band 6 & $0''.048 \times 0''.030$ &$<4.1$ & $\sim1.37$ \\
HD169142 & Band 6 & $0''.072 \times 0''.043$ & $<4.9$& $\sim0.0076$ \\
HD135344B & Band 6 & $0''.245 \times 0''.180$ & $<24.5$ & $\sim0.013$\\
MWC758 & Band 7 & $0''.040 \times 0''.031$ & $<5.0$ & $\sim0.0042$\\
SR24S & Band 6 & $0''.043 \times 0''.035$ & $<4.0$ & $\sim0.022$\\
TCha & Band 3 & $0''.296 \times 0''.170$ & $<18.7$ &$\sim0.15$\\

\hline

\end{tabular}
\end{center}
\tablefoot{Columns show the target name, the band which flux has been used for the mass estimate, and the estimated inner disk dust mass in Earth mass. {The dust masses here reported are highly uncertain. See the text for a more complete discussion.}}

\end{table*}

\subsection{Photoevaporative wind or jet-and-MHD wind}

\begin{figure} 
   \centering
   \includegraphics[width=0.5\textwidth]{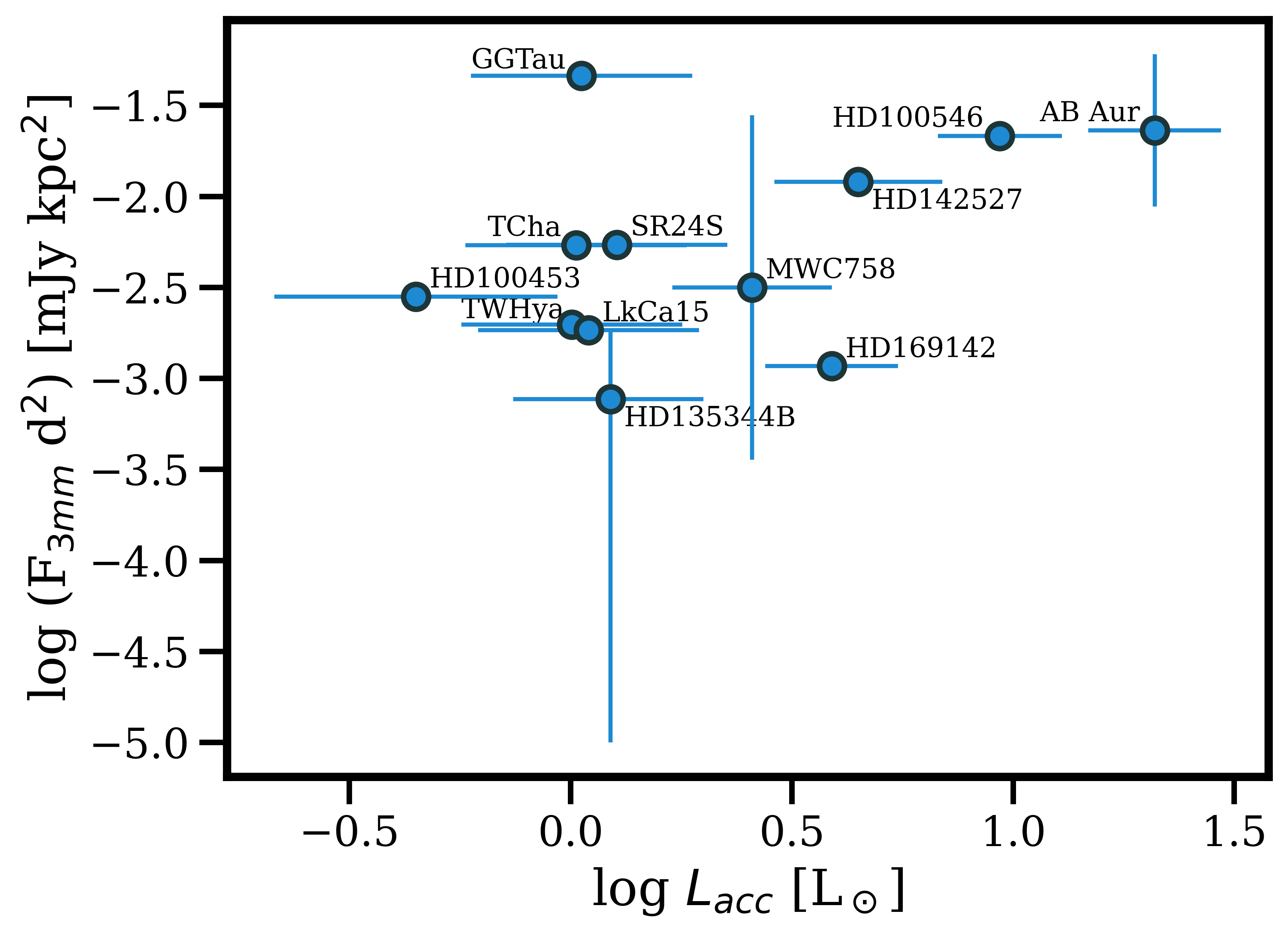}
   \centering
   \includegraphics[width=0.5\textwidth]{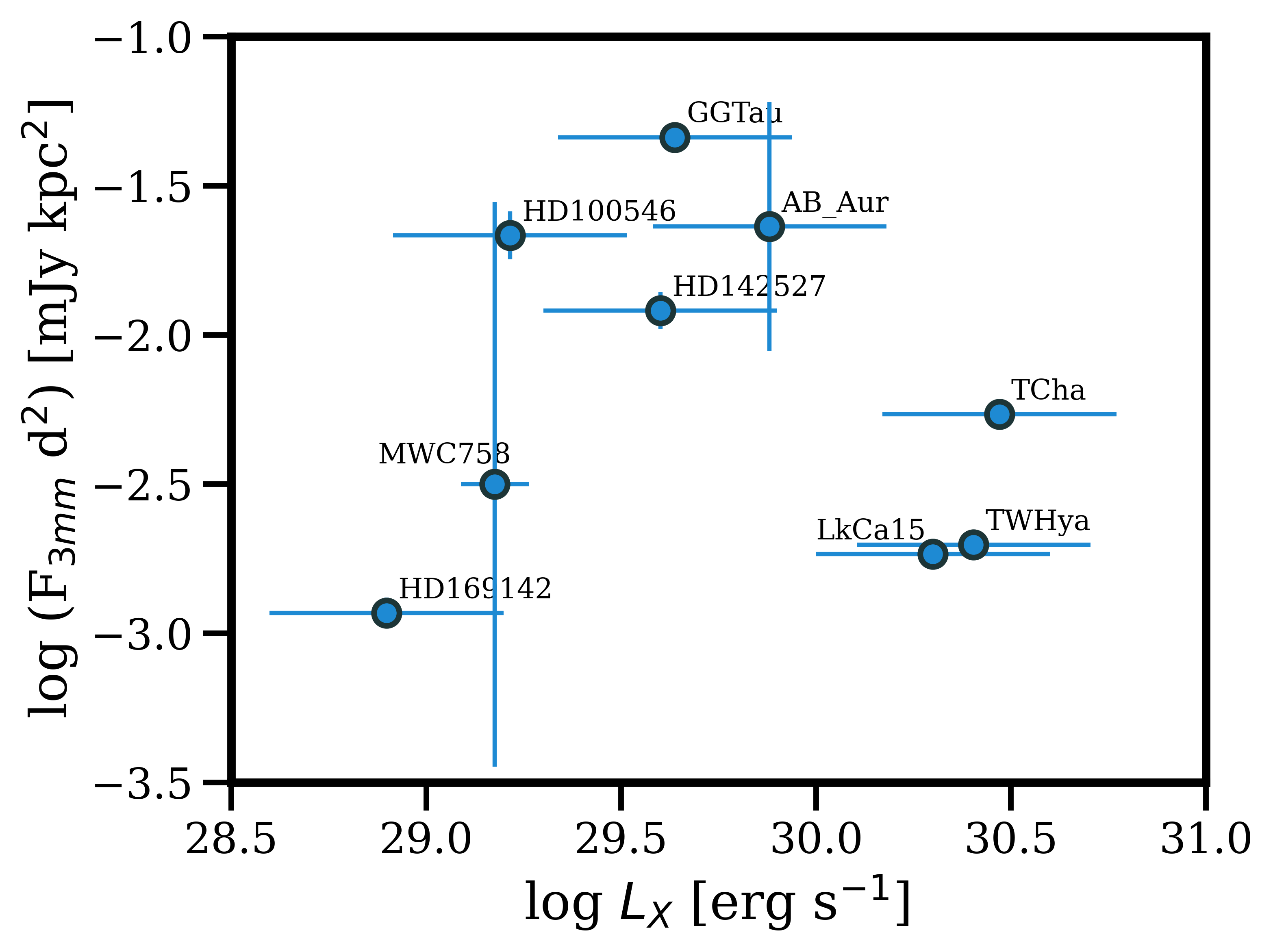}
      \caption{Observed luminosity at 3 mm as a function of the accretion luminosity onto the central star (\textit{top}) and as a function of the X-ray luminosity onto the central star (\textit{bottom}).
      {Since ALMA Band 3 data were not available, the luminosity at 90 GHz for MWC758, AB Aur, and HD135344B are extrapolated using the estimated spectral index, from either Bands 7 or 4 data.}
      In all other cases, Band 3 luminosities are shown. The X-ray luminosities of HD100453, HD135344B, and SR24S are not available in the literature (and, thus, not shown). }
    \label{fig:fluxVSLaccX}  
\end{figure}

The central emission is thus dominated by free-free emission from either a jet or a photoevaporative wind. With the assumption that the central emission is associated with a photoevaporative wind, in the optically thin domain, the free-free luminosity is linearly dependent on the extreme-ultraviolet (EUV) luminosity $\Phi_\text{EUV}$ -- in turn expected to be correlated with the accretion luminosity, $L_\text{acc}$ (\citealp{2016ErcolanoOwen}) -- or on the X-ray luminosity, $L_X$, in the case of fully or partially ionized wind, respectively (\citealp{2012Pascucci}, \citealp{2014Pascucci}). 
The top panel in Figure \ref{fig:fluxVSLaccX} shows that no strong correlation between the free-free luminosity, as inferred from the millimeter flux, and the accretion luminosity is found (pearson correlation coefficient $r=0.49\pm0.28$, p-value $p=0.10$). Moreover, as shown by the bottom panel in the figure, no correlation between the free-free luminosity and the X-ray luminosity is found (pearson correlation coefficient $r=-0.09\pm0.37$, p-value $p=0.82$). 
The central free-free emission is thus not likely to be associated with a photoevaporative wind. {Moreover, flat or positive spectral indices, as the one estimated for our sample, are expected to be associated with partially optically thick free-free emission due to other mechanisms, such as collimated ionized outflows and jets, rather than photoevaporative winds (\citealp{2012Pascucci}).}

\subsection{Properties of the jet emission}

\begin{figure} 
   \centering
   \includegraphics[width=0.5\textwidth]{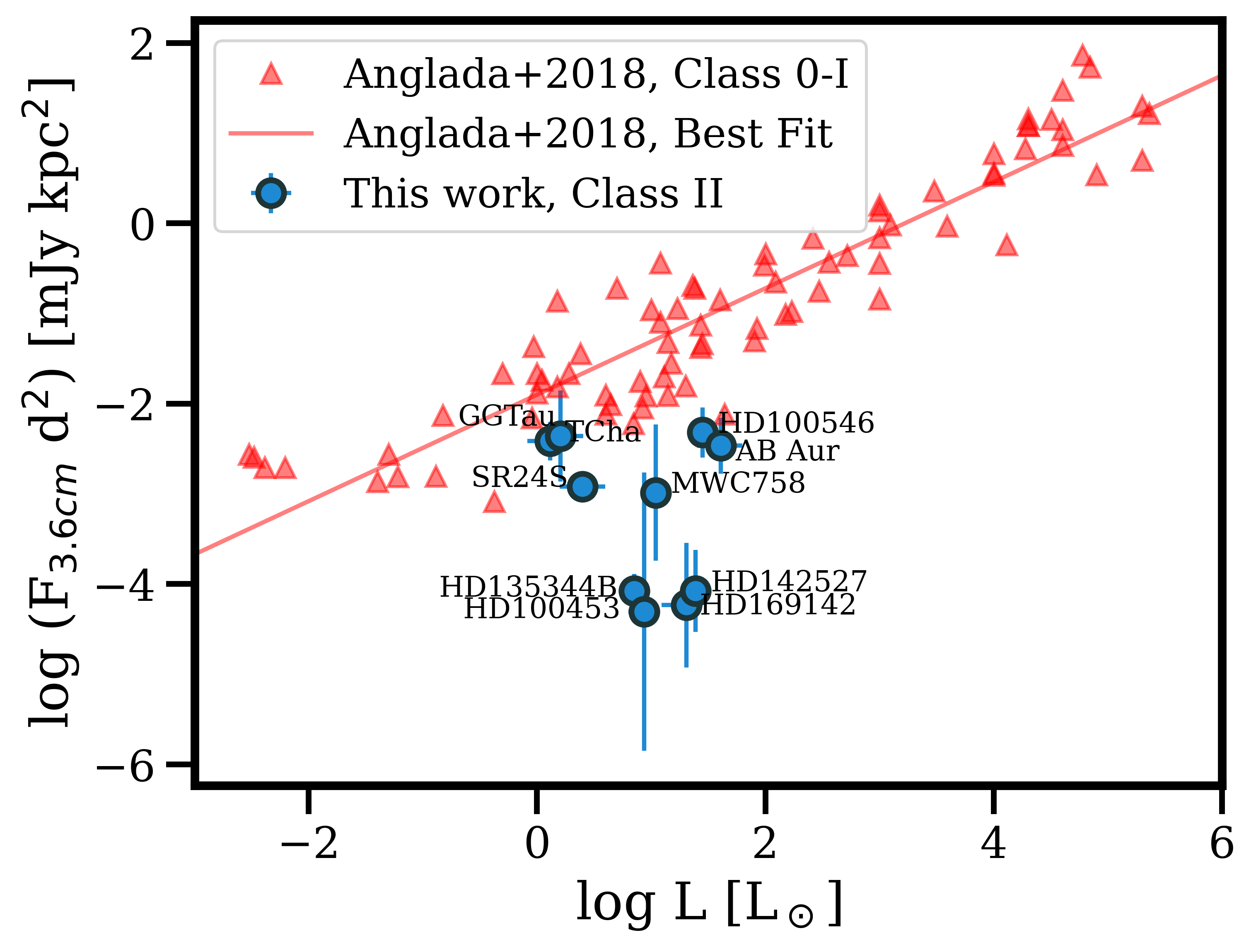}
      \caption{3.6 cm luminosity as a function of the stellar/bolometric luminosity for the central emission observed in our sample and for Class 0-I radio jets (\citealp{2018Anglada}). The disk around WSB60 is not shown since the emission is dominated by thermal dust contribution.}
    \label{fig:fluxVSlum}  
\end{figure}

Assuming that the central emission is associated with a jet, we compared the observed fluxes in our sample of Class II disks to the fluxes of radio jets observed at 3.6 cm in a sample of younger Class 0-I disks (\citealp{2018Anglada}). Figure \ref{fig:fluxVSlum} shows this comparison. The blue dots show the expected luminosities at 3.6 cm in our sample, extrapolated using Equation \ref{eq:corrF} and the estimated $\alpha_\text{inner}$ (Table \ref{tab:alpha}), while the red triangles show the luminosities of the radio jets, reported in \citealp{2018Anglada}.  
The Class II disks in our sample are found to be underluminous with respect to the Class 0-I disks. As discussed in \cite{2018Anglada}, the accretion component of the luminosity is correlated with the radio flux of the jet, and thus with the outflow.
While the accretion luminosity is dominant in the youngest
objects, in more evolved sources the total luminosity is more affected by the stellar contribution (which is not expected to be correlated with the radio emission); thus, it is expected that in Class II disks, as the ones studied in this work, the 3.6 cm luminosity is correlated with only a fraction of the bolometric luminosity.

We now investigate the relation between the free-free emission and the accretion rate onto the central star. Under the assumption that the central emission in those ten systems is associated with the gas from an ionized jet, we estimated the ionized mass loss rate $\dot M_\text{i}$ following the modeling of the free-free emission from a jet by \cite{1986Reynolds} (see also \cite{2018Anglada}):

\begin{equation}\label{eq:Mion}
\begin{split}
    \left(\frac{\dot M_\text{i}}{10^{-6} M_\odot \text{yr}^{-1}} \right) & = 0.108 \left[   \frac{(2-\alpha)(0.1+\alpha)}{1.3-\alpha} \right]^{0.75} \\
  & \times \left[ \left(\frac{F_\nu}{\text{mJy}}\right)\left(\frac{\nu}{10 \text{GHz}}\right)^{-\alpha} \right]^{0.75} \left( \frac{v_\text{jet}}{200 \text{km s}^{-1}} \right)  \\
  & \times  \left( \frac{\nu_\text{m}}{10 \text{GHz}} \right)^{0.75\alpha - 0.45} \left( \frac{\theta_0}{\text{rad}}\right)^{0.75} (\sin i)^{-0.25} \\
  & \times \left( \frac{d}{\text{kpc}} \right)^{1.5} \left( \frac{T}{10^4 \text{K}}\right)^{-0.075},
\end{split}
\end{equation}
where $\alpha$ is the spectral index, $F_\nu$ is the continuum flux at the frequency $\nu$ (observed or extrapolated at 2-3 mm in our case, or 1 mm when the emission is not detected at 2-3 mm), and $d$ is the distance to the source. The injection opening angle of the jet $\theta_0$ is approximated as $2 \arctan(\theta_\text{min}/\theta_\text{maj})$ (\citealp{2018Anglada}), with the ratio between the minor and major axis of the jet $\theta_\text{min}/\theta_\text{maj}$ assumed equal to 0.5 (see \citealp{2021Kavak}). A value of $T=10^4$ K is adopted for the ionized gas (see \citealp{2018Anglada}). 
{Since the value of the turnover frequency has not been determined directly from observations yet (\citealp{2018Anglada}), a turnover frequency of $\nu=40$ GHz is assumed (\citealp{2021Kavak}). However, since the dependence of the ionized mass loss rate on $\nu$ is almost negligible and disappears for $\alpha = 0.6$, this assumption will not significantly impact the estimate of $\dot M_\text{i}$.}

We assumed that the jet is perpendicular to the plane of the disk, and thus the inclination of the outer disk (reported in Table \ref{tab:sample}) is taken as the inclination of the jet $i$. 
Finally, we assumed that the velocity of the jet $v_\text{jet}$ is (see the discussion in \citealp{2018Anglada}): 
\begin{equation}
    \left( \frac{v_\text{jet}}{\text{km s}^{-1}} \right) \simeq 140 \left( \frac{M_*}{0.5 M_\odot} \right)^{1/2}.
\end{equation}

\begin{figure} 
   \centering
   \includegraphics[width=0.5\textwidth]{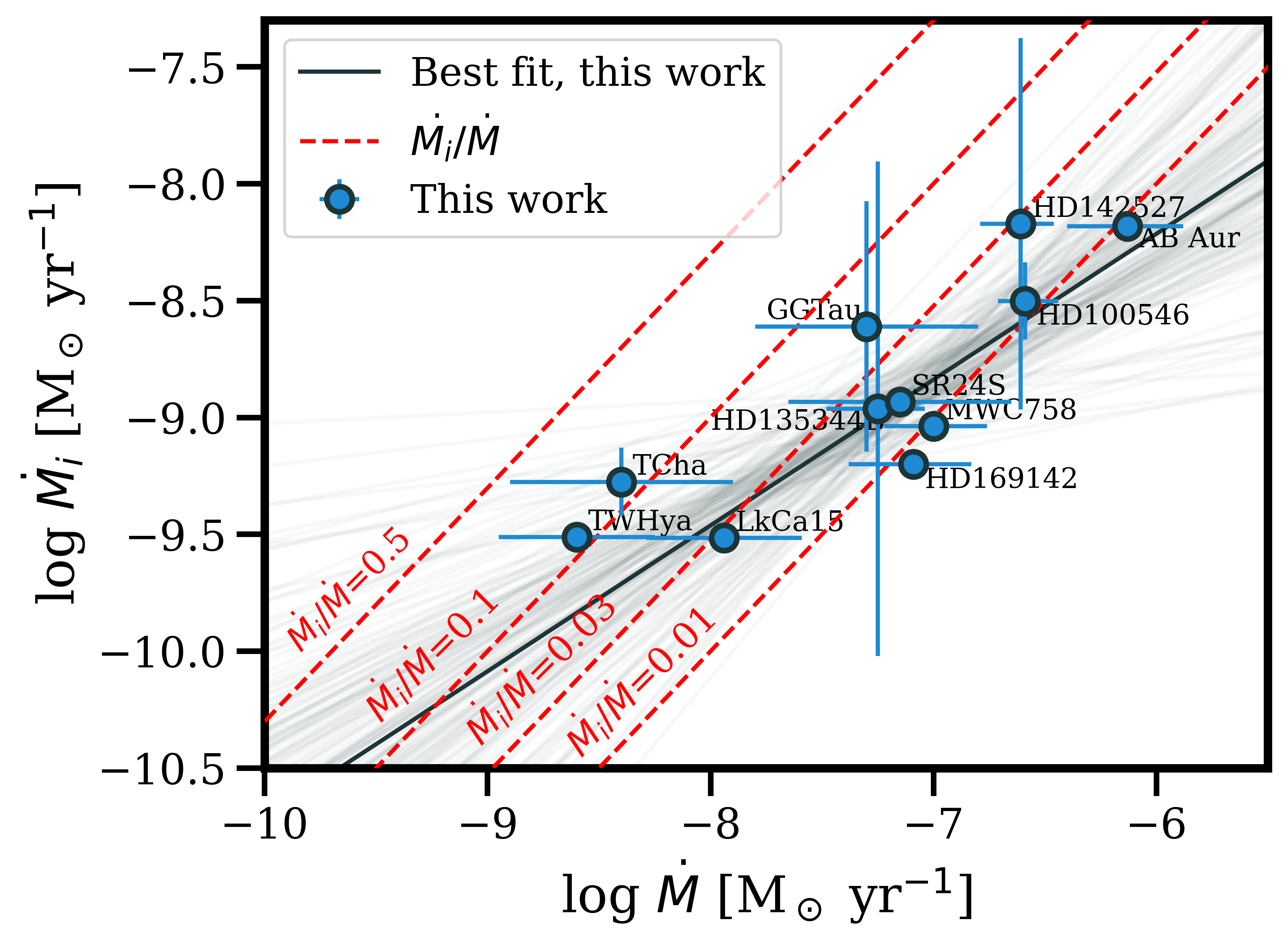}
      \caption{Ionized mass loss rate as a function of the accretion rate onto the central star for the central emission observed in the sample. The black line shows the best-fit correlation.}
    \label{fig:MionVSMacc}  
\end{figure}
\begin{figure} 
   \centering
   \includegraphics[width=0.5\textwidth]{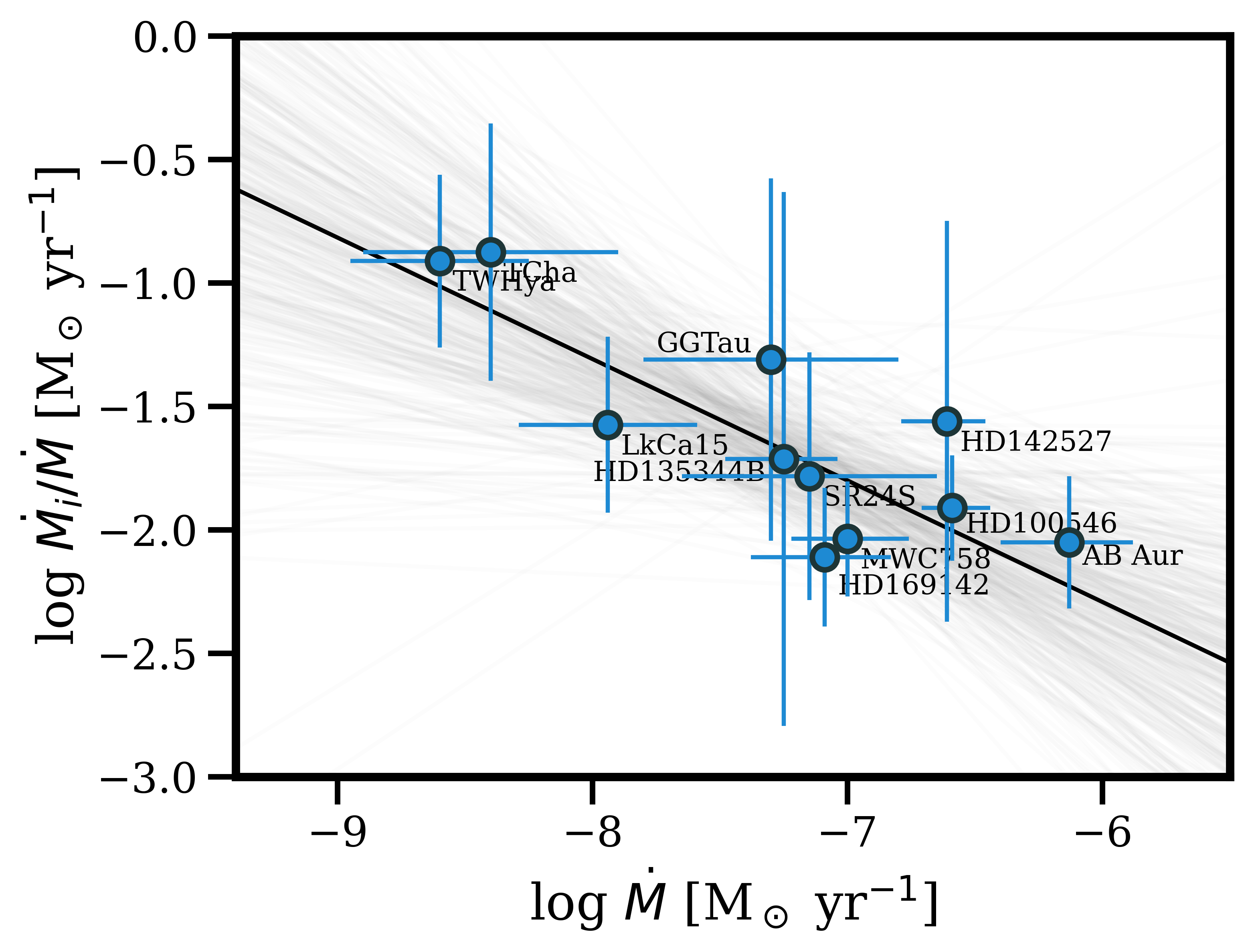}
      \caption{Jet efficiency in regulating the mass transfer from the disk to the star, $\dot M_{i}/\dot M$, as a function of the accretion rate onto the central star. The black line (and gray lines) shows the best-fit correlation (with relative uncertainty).}
    \label{fig:ratioVSMacc}  
\end{figure}

The ionized mass loss rate cannot be computed for the disk around HD100453 and WSB60, where, since the dust is probably significantly contributing to the emission, the estimated spectral index is $>1.3$ and the denominator in the first line in Equation \ref{eq:Mion} is negative. Table \ref{tab:alpha} reports the values for $\dot M_\text{i}$ calculated for the disks in our sample. 
In addition to the disks in our sample, we computed the ionized mass loss rate for the central emission in TW Hya and LkCa 15 (\citealp{2021Macias}, \citealp{2020Facchini}). As for TW Hya, \cite{2021Macias} reported a the spectral index of the central emission of $\alpha_\text{inner}=0.65\pm0.05$, while the accretion rate onto the central star is $\log \dot M = -8.60 \pm 0.35 M_\odot / \text{yr}$ (\citealp{2019AVenuti}). The spectral index of LkCa 15, reported by \cite{2020Facchini}, is $\alpha_\text{inner}=0.9\pm0.2$, with an accretion rate of $\log \dot M = -7.94 \pm 0.35 M_\odot / \text{yr}$ (\citealp{2023ManaraPPVII}). 
Figure \ref{fig:MionVSMacc} shows the ionized mass loss rate estimated from the detected free-free emission flux as a function of the accretion rate, as determined from H$\alpha$ emission (see Table \ref{tab:sample}). 
The ionized mass loss rates span from $\sim 10^{-10}$ to $\sim 10^{-8}  M_\odot / \text{yr}$ and the corresponding $\dot M_{i}/\dot M$ ratios range between $\sim 0.01$ and $\sim 0.1$, with a median ratio of {0.019}.
A correlation between mentioned properties is found with a pearson correlation coefficient, $r =0.86 \pm 0.17$ and p-value $p=0.0009$, and with the following relationship\footnote{The liner regression was performed with \texttt{linmix} (\citealp{2007Kelly})}:
\begin{equation}
   \left( \frac{\dot M_{i}}{M_\odot \text{\text{yr}}^{-1}} \right) = 10^{-4.47 \pm 1.53} \left(\frac{\dot M}{\text{M}_\odot \text{yr}^{-1}} \right)^{0.62\pm 0.21}.
\end{equation}
This positive correlation between the accretion rate and the ionized mass loss rate, as inferred from the free-free emission, implies that {jets and/or ionized MHD winds are one of the main drivers of accretion in protoplanetary disks}. 
Moreover, the ratio $\dot M_{i}/\dot M$ characterizes the jet efficiency $\xi$ of the outflow launch, defined as $   \xi = (1/\dot M) (\text{d}\dot M_i/\text{d}\log R) \sim (\dot M_i / \dot M) (R /\text{d}R)$ (e.g., \citealp{2023Lesurppvii}, \citealp{2023PascucciPPVII}).  
Since the correlation is slightly sub-linear, the jet efficiency, and thus the efficiency of the angular momentum extraction, is anti-correlated with the stellar accretion rate (Figure \ref{fig:ratioVSMacc}). Similar behavior can commonly be seen in MHD simulations (e.g., \citealp{2021Lesur}), where higher magnetizations lead to higher accretion rates but lower efficiencies.

These results are consistent with ionized mass ejection rates measured from HVC [O I]6300 $\mathring{\mathrm{A}}$ line luminosity in a sample of 131 young stars with disks in the Lupus, Chamaeleon and $\sigma$Orionis star-forming region (\citealp{2018Nisini}). The $\dot M_{i}/\dot M$ ratio in that sample ranges from $\sim 0.01$ to $\sim0.5$,  with the tendency for sources with higher accretion rates ($>10^{-8}$M$_\odot$/yr, as those analyzed in this work) of displaying ratios lower than 0.1 and with 44\% of the entire sample (considering both detections and upper limits), having $\dot M_{i}/\dot M <0.03$.
Although the free-free emission is known to be highly variable (e.g., \citealp{2017Ubach}), its variability has not been well characterized yet and its origin is still unclear. On the other hand, the accretion rate onto the central star has been shown to be variable with time in some cases (e.g., \citealp{2014Venuti}). The observed correlation between the accretion rate and the ionized mass loss rate, as inferred from the free-free emission, shows that the variability of these two properties might be as well correlated. {Nevertheless, \citealp{2022Espaillat} report no variability in the 3 cm emission of GM Aur observed in a monitoring campaign with VLA, concluding that changes of less than a factor of $\sim3$ in the accretion rates do not lead to detectable changes in the mass-loss rate traced by the jet at cm-wavelengths.}
Moreover, since a correlation is observed despite the accretion rates and the free-free emission having not been estimated at the same time, the impact of variability of the emission on the $\dot M_{i}/\dot M$ should be small. Therefore, the observed scatter in the $\dot M_{i}/\dot M$ may not reflect a real difference in the jet efficiency among sources. 
A study on a larger sample of disks with a long-term monitoring of these properties is needed to better understand and characterize the interplay between the accretion rate and the free-free emission.

In summary, the positive correlation between the ionized mass loss rate and the accretion rate found in this work confirm the connection between accretion and outflow, already expected from studies of the hot neutral gas traced by the HVC [O I]6300 $\mathring{\mathrm{A}}$ line, with an independent new method, opening up new possibilities in analyzing disk evolution and angular momentum transport processes, even in targets where HVC [O I]6300 $\mathring{\mathrm{A}}$ are usually not detected.



\section{Conclusions}\label{conclusions}
In this work, we analyze archival ALMA observations at 1 mm and 3 mm of 11 transition disks in nearby star-forming regions. Central emission close to the star has been detected at multiple frequencies in all targets in our sample. The analysis of the spectral energy distribution showed that the millimeter spectral index of the emission of the outer disk is consistent with dust growth, similar to regular protoplanetary disks. On the other hand, the central emission close to the star is likely associated with free-free emission from an ionized jet or a disk wind ($0 < \alpha < 1$) in all disks in the sample, except WSB60. We summarize our main conclusions below:
\begin{itemize}
    \item The lack of mm-dust grains in the majority of inner disks in transition disks suggests that either such dust grains have drifted quickly towards the central star, grain growth is less efficient in the inner disk, or grains grow rapidly to planetesimal sizes in the inner disk.
    
\item The central cm-wave luminosity of the transition disks at a given bolometric luminosity is found to be underluminous with respect to younger Class 0-I objects, indicating that the accretion component is {a small fraction of their bolometric luminosity}.

\item No correlation between the free-free luminosity and the accretion luminosity  or the X-ray luminosity was found, suggesting that the central free-free emission in these disks is more likely associated with a jet{ and/or MHD wind,} rather than with a photoevaporative wind.

\item Under the assumption that the central emission is associated with a jet, a correlation with a pearson coefficient $r=0.89\pm0.15$ and p-value $p=0.0002$ is found between the ionized mass loss rate and the accretion rate onto the central star, confirming that the outflow is strictly connected with the stellar accretion, implying that accretion in these disks is {mainly} driven by a jet. 
This correlation is slightly sub-linear, pointing out that the jet efficiency in regulating the mass transfer from the disk to the star is anti-correlated with the stellar accretion rate.

\end{itemize}

Observations in a larger sample of transition disks, including centimeter and ALMA Band 1 data, where the contribution of the free-free emission is expected to dominate, will be crucial to minimize the contribution of the thermal dust to the spectral index estimate and to confirm that the central emission is associated with free-free emission. Moreover, the estimate of the spectral index of the central emission in a larger sample will allow for  the observed correlation between the accretion rate and the ionized mass loss rate to be strengthened. Finally, a long-term monitoring of the accretion rate and of the ionized mass loss rate is needed to better characterize the interplay between these two properties and to better constrain the impact of the variability of the emission on the observed correlation.

\begin{acknowledgements}
{The authors thank the referee for their useful comments that have contributed to improve the manuscript. AAR thanks Andrea Isella for useful discussion.}
This paper makes use of the following ALMA data: 2012.1.00158.S, 2012.1.00182.S, 2012.1.00303.S, 2012.1.00799.S, 2013.1.00157.S, 2013.1.00670.S, 2015.1.00425.S, 2015.1.00806.S, 2015.1.00889.S, 2015.1.00979.S, 2015.1.01353.S, 2016.1.00340.S, 2016.1.00344.S, 2016.1.01042.S, 2017.1.00492.S, 2017.1.00884.S, 2017.1.00885.S, 2017.1.00987.S, 2017.1.01419.S, 2017.1.01678.S, 2018.1.00028.S, 2018.1.00532.S, 2018.1.00618.S, 2018.1.01309.S, 2018.1.01716.S, 2019.1.00579.S. ALMA is a partnership of ESO (representing its member states), NSF (USA) and NINS (Japan), together with NRC (Canada), MOST and ASIAA (Taiwan), and KASI (Republic of Korea), in cooperation with the Republic of Chile. The Joint ALMA Observatory is operated by ESO, AUI/NRAO and NAOJ. The PI acknowledges assistance from Allegro, the European ALMA Regional Center node in the Netherlands.     
\end{acknowledgements}

\bibliography{bibliography.bib}
\appendix

\section{Intensity maps}\label{app:propImages}

\begin{figure*} 
   \centering
   \includegraphics[width=0.5\textwidth]{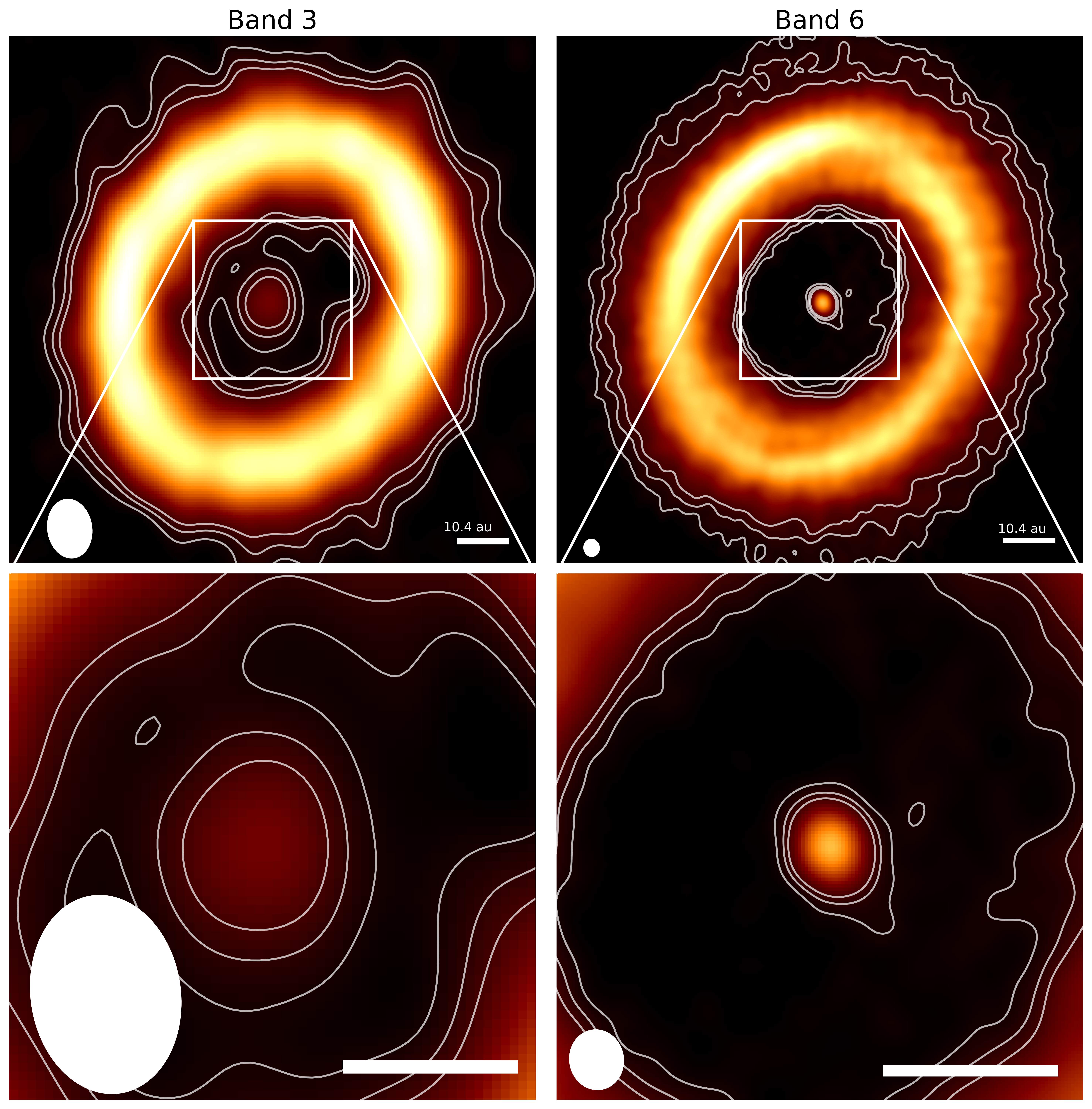}
      \caption{Intensity maps of the disk around HD100453. The first and second columns show the images of the disk in Band 3 and Band 6, respectively. Band 7 observations used in this work are from \citealp{2020Rosotti}. The images in the first row are $1''.0 \times 1''.0$, while the zooms in the second row are $0''.3 \times 0''.3$.  In each image, the color scale has the peak flux as the maximum, and the image rms as minimum. All bars in the bottom right are $0''.1$ in length, which is $\sim10.4$ au at the distance of the source. White contours show three, five, and seven times the rms of the continuum emission.}
    \label{fig:HD100453}  
\end{figure*}

\begin{figure*} 
   \centering
   \includegraphics[width=0.7\textwidth]{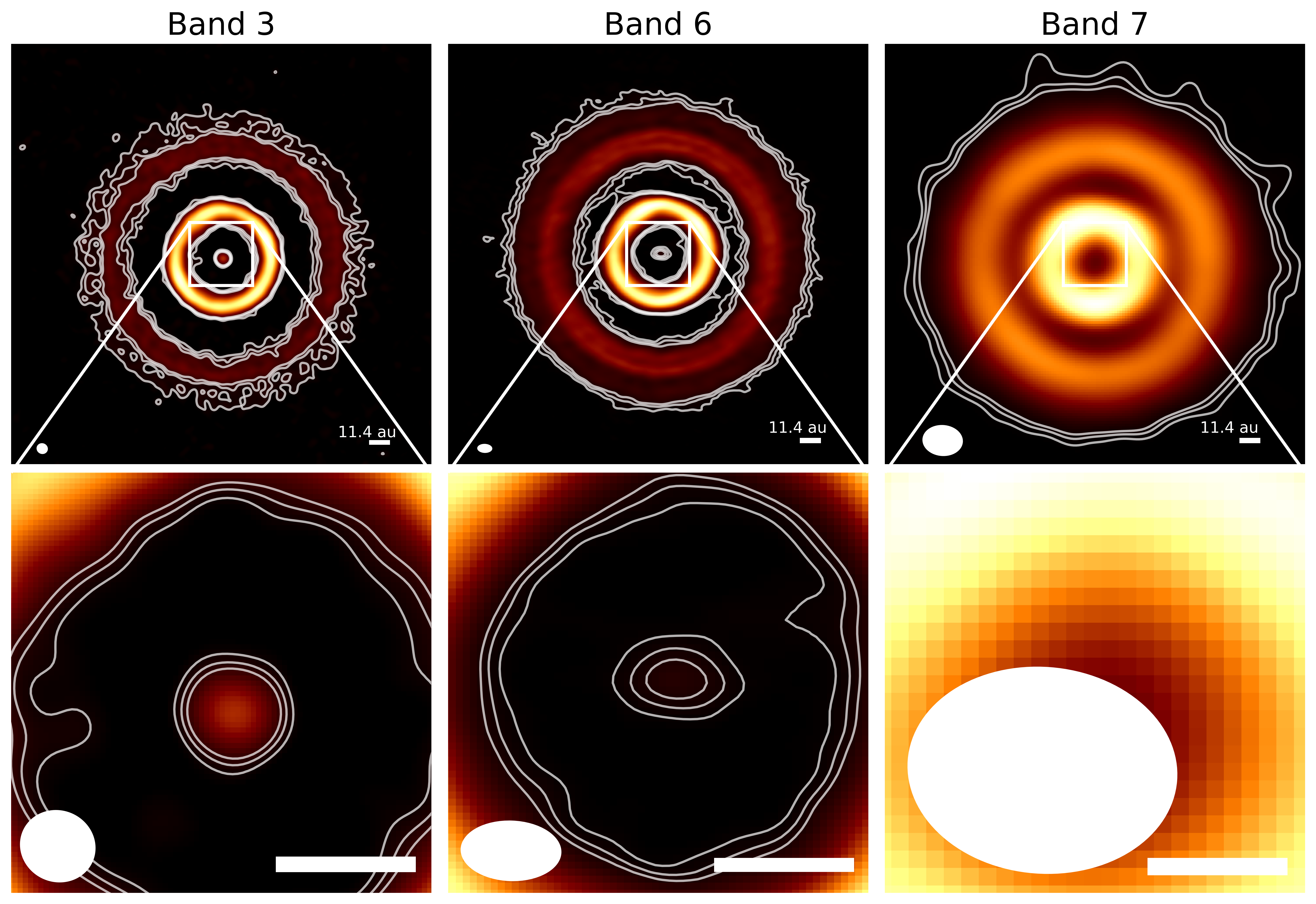}
      \caption{Intensity maps of the disk around HD169142. The first, second, and third columns show the images of the disk in Band 3, Band 7, and Band 7, respectively. The images in the first row are $2''.0 \times 2''.0$, while the zooms in the second row are $0''.3 \times 0''.3$.  In each image, the color scale has the peak flux as the maximum, and the image rms as minimum. All bars in the bottom right are $0''.1$ in length, which is $\sim11.4$ au at the distance of the source. White contours show three, five, and seven times the rms of the continuum emission.}
    \label{fig:HD169142}  
\end{figure*}

\begin{figure*} 
   \centering
   \includegraphics[width=0.7\textwidth]{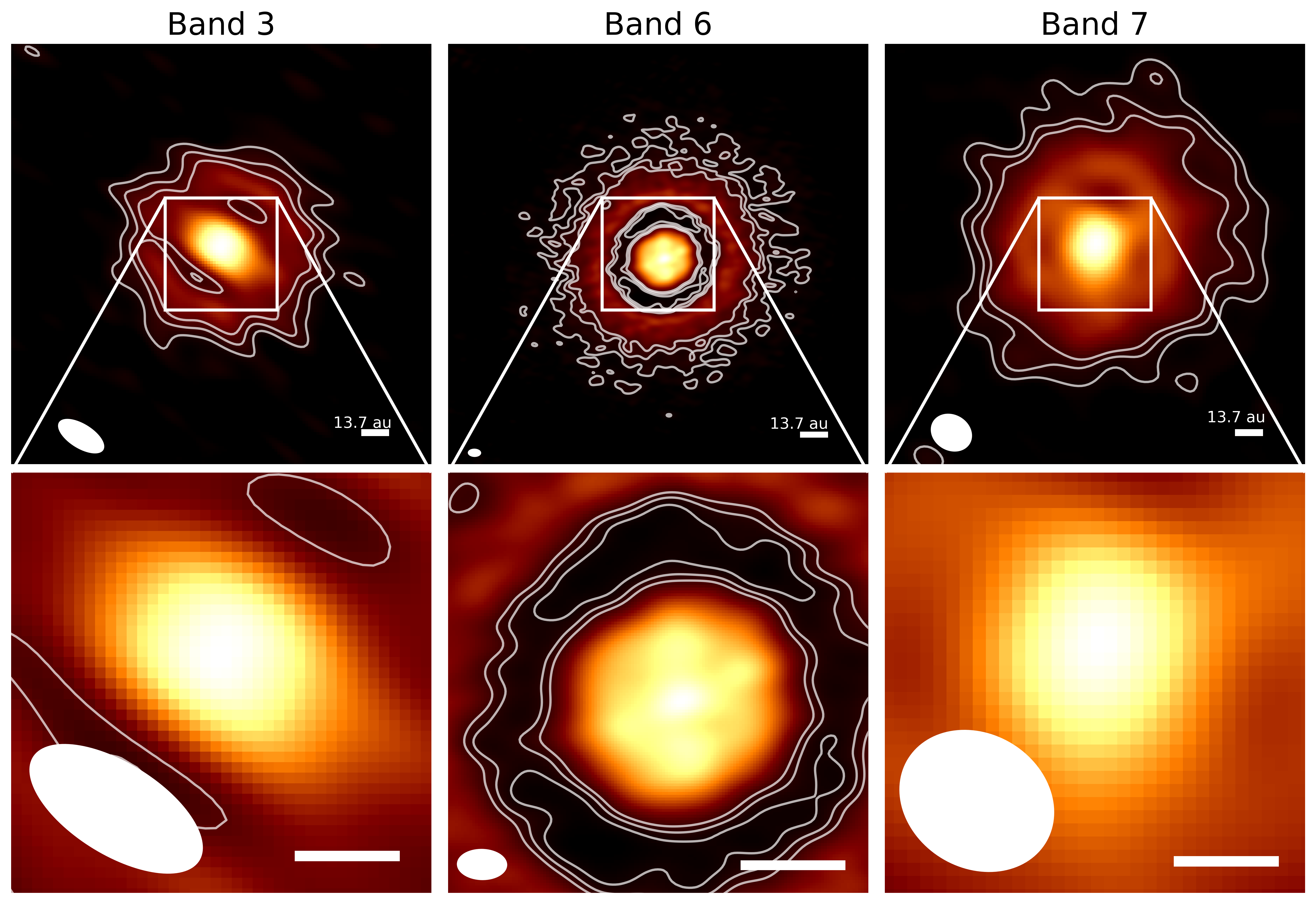}
      \caption{Intensity maps of the disk around WSB60. The first, second, and third columns show the images of the disk in Band 3, Band 7, and Band 7, respectively. The images in the first row are $1''.5 \times 1''.5$, while the zooms in the second row are $0''.4 \times 0''.4$.  In each image, the color scale has the peak flux as the maximum, and the image rms as minimum. All bars in the bottom right are $0''.1$ in length, which is $\sim13.7$ au at the distance of the source. White contours show three, five, and seven times the rms of the continuum emission.}
    \label{fig:wsb60}  
\end{figure*}

\begin{figure*} 
   \centering
   \includegraphics[width=0.5\textwidth]{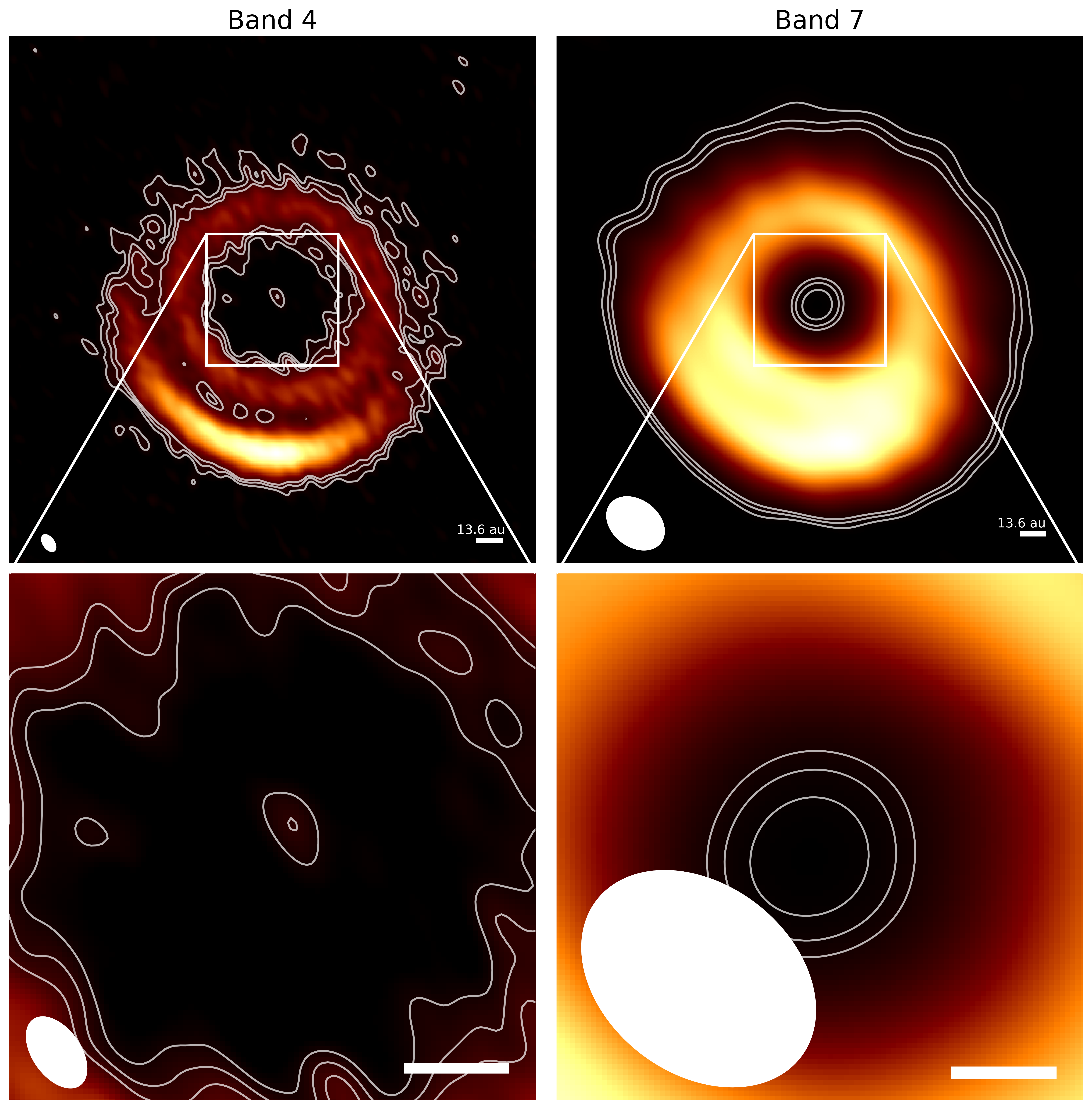}
      \caption{Intensity maps of the disk around HD1035344B. The first and second columns show the images of the disk in Band 4 and Band 6, respectively. Band 6 observations used in this work are from \citealp{2015Casassus}. The images in the first row are $2''.0 \times 2''.0$, while the zooms in the second row are $0''.5 \times 0''.5$.  In each image, the color scale has the peak flux as the maximum, and the image rms as minimum. All bars in the bottom right are $0''.1$ in length, which is $\sim13.6$ au at the distance of the source. White contours show three, five, and seven times the rms of the continuum emission.}
    \label{fig:HD135344b}  
\end{figure*}

\begin{figure*} 
   \centering
   \includegraphics[width=0.7\textwidth]{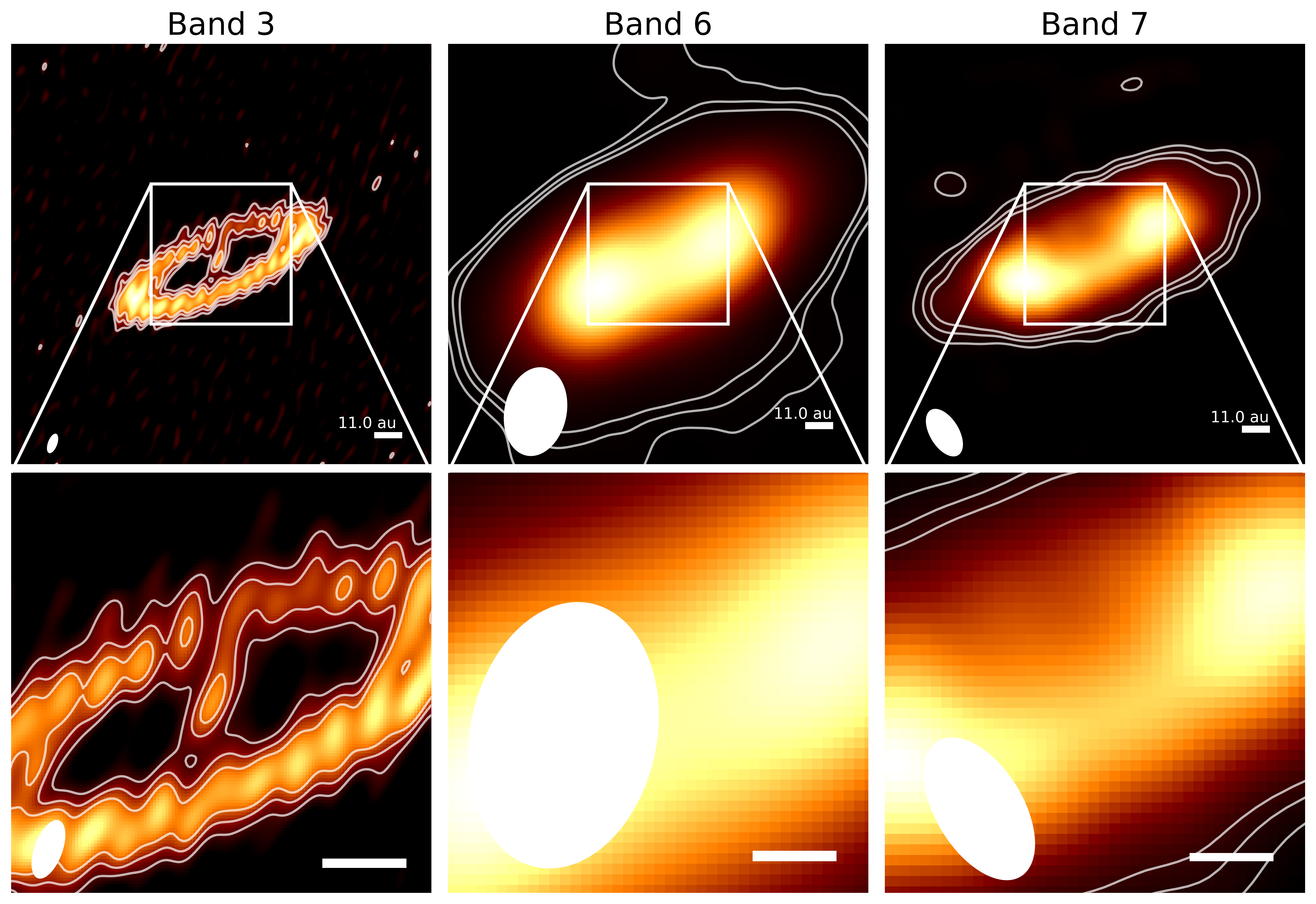}
      \caption{Intensity maps of the disk around T Cha. The first, second, and third columns show the images of the disk in Band 3, Band 6, and Band 7, 
      respectively. The images in the first row are $1''.5 \times 1''.5$, while the zooms in the second row are $0''.3 \times 0''.3$.  In each image, the color scale has the peak flux as the maximum and the image rms as minimum. All bars in the bottom right are $0''.1$ in length, which is $\sim11.0$ au at the distance of the source. White contours show three, five, and seven times the rms of the continuum emission.}
    \label{fig:TCha}  
\end{figure*}

\begin{figure*} 
   \centering
   \includegraphics[width=0.7\textwidth]{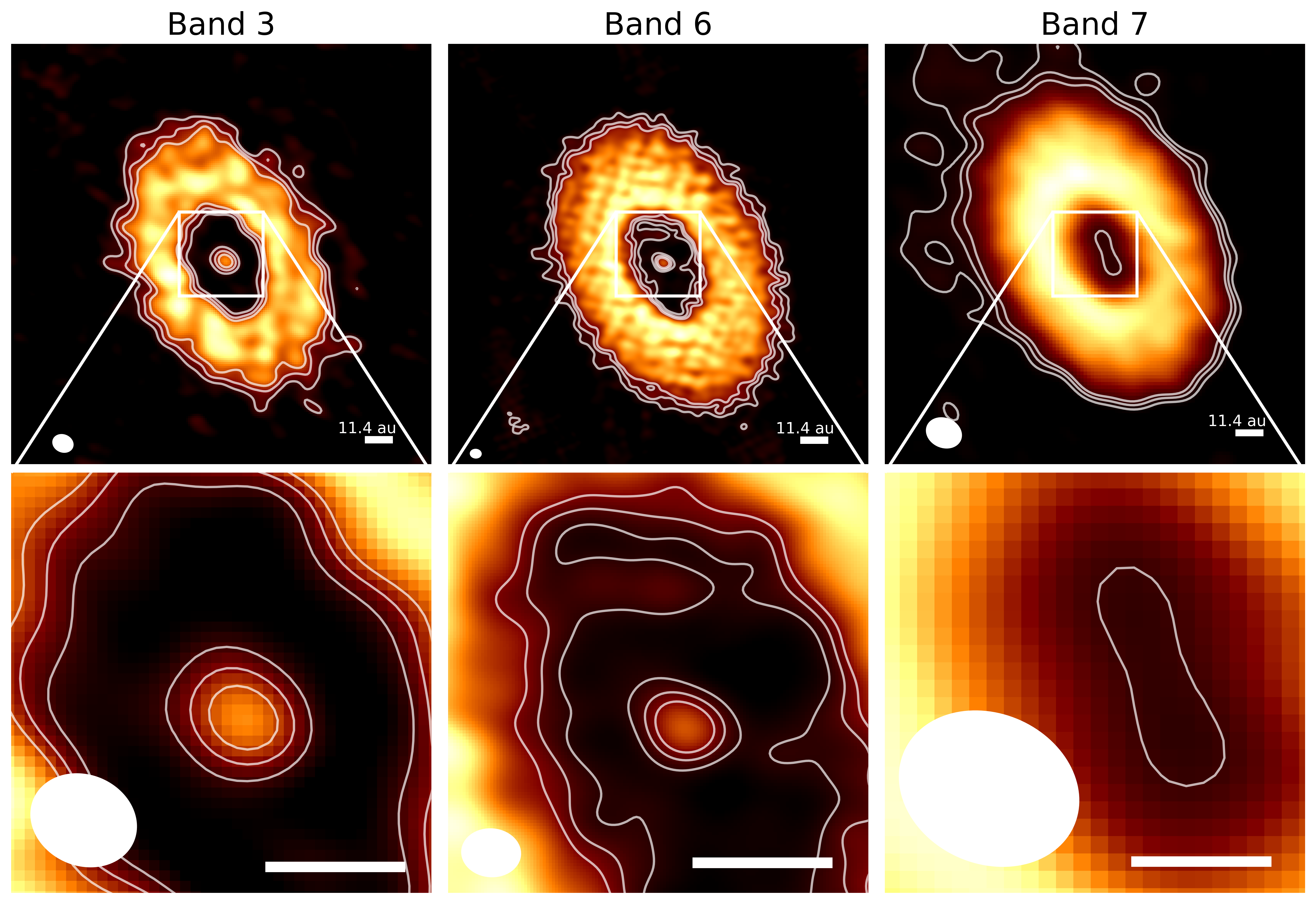}
      \caption{Intensity maps of the disk around SR24S. The first, second, and third columns show the images of the disk in Band 3, Band 6, and Band 7, respectively. The images in the first row are $1''.5 \times 1''.5$, while the zooms in the second row are $0''.5 \times 0''.5$.  In each image, the color scale has the peak flux as the maximum, and the image rms as minimum. All bars in the bottom right are $0''.1$ in length, which is $\sim11.4$ au at the distance of the source. White contours show three, five, and seven times the rms of the continuum emission.}
    \label{fig:SR24S}  
\end{figure*}

\begin{figure*} 
   \centering
      \includegraphics[width=0.8\textwidth]{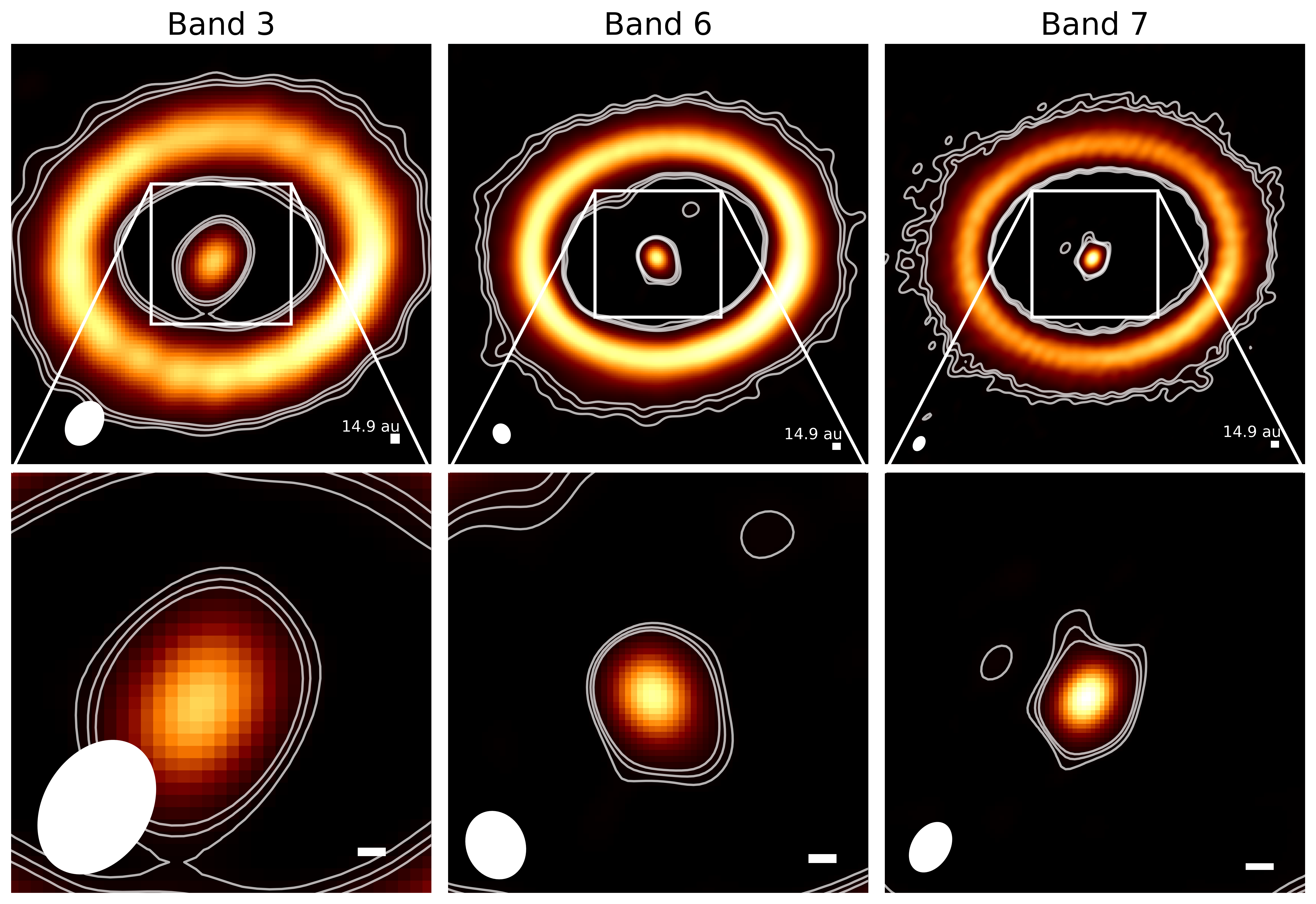}
      \caption{Intensity maps of the disk around GG Tau. The first, second, and third columns show the images of the disk in Band 3, Band 6, and Band 7, respectively. The images in the first row are $5''.0 \times 5''.0$, while the zooms in the second row are $1''.5 \times 1''.5$.  In each image, the color scale has the peak flux as the maximum, and the image rms as minimum. All bars in the bottom right are $0''.1$ in length, which is $\sim14.9$ au at the distance of the source. White contours show three, five, and seven times the rms of the continuum emission.}
    \label{fig:GGTau}  
\end{figure*}

\begin{figure*} 
   \centering
   \includegraphics[width=0.8\textwidth]{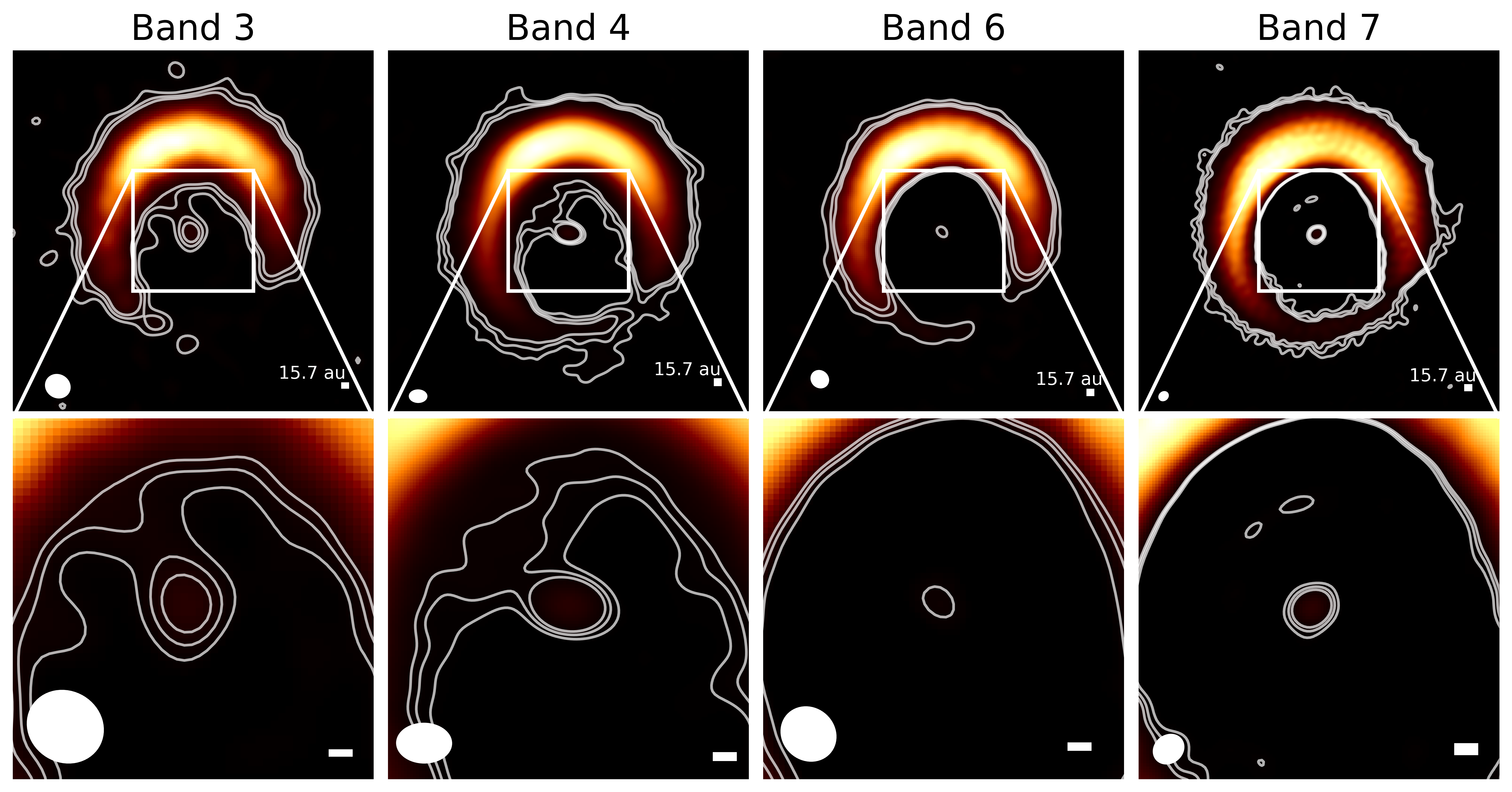}
      \caption{Intensity maps of the disk around HD142527. The first, second, third, and fourth columns show the images of the disk in Band 3, Band 4, Band 6, and Band 7, respectively. The images in the first row are $4''.5 \times 4''.5$, while the zooms in the second row are $1''.5 \times 1''.5$.  In each image, the color scale has the peak flux as the maximum, and the image rms as minimum. All bars in the bottom right are $0''.1$ in length, which is $\sim15.7$ au at the distance of the source. White contours show three, five, and seven times the rms of the continuum emission.}
    \label{fig:HD142527}  
\end{figure*}

\begin{figure*} 
   \centering
   \includegraphics[width=0.3\textwidth]{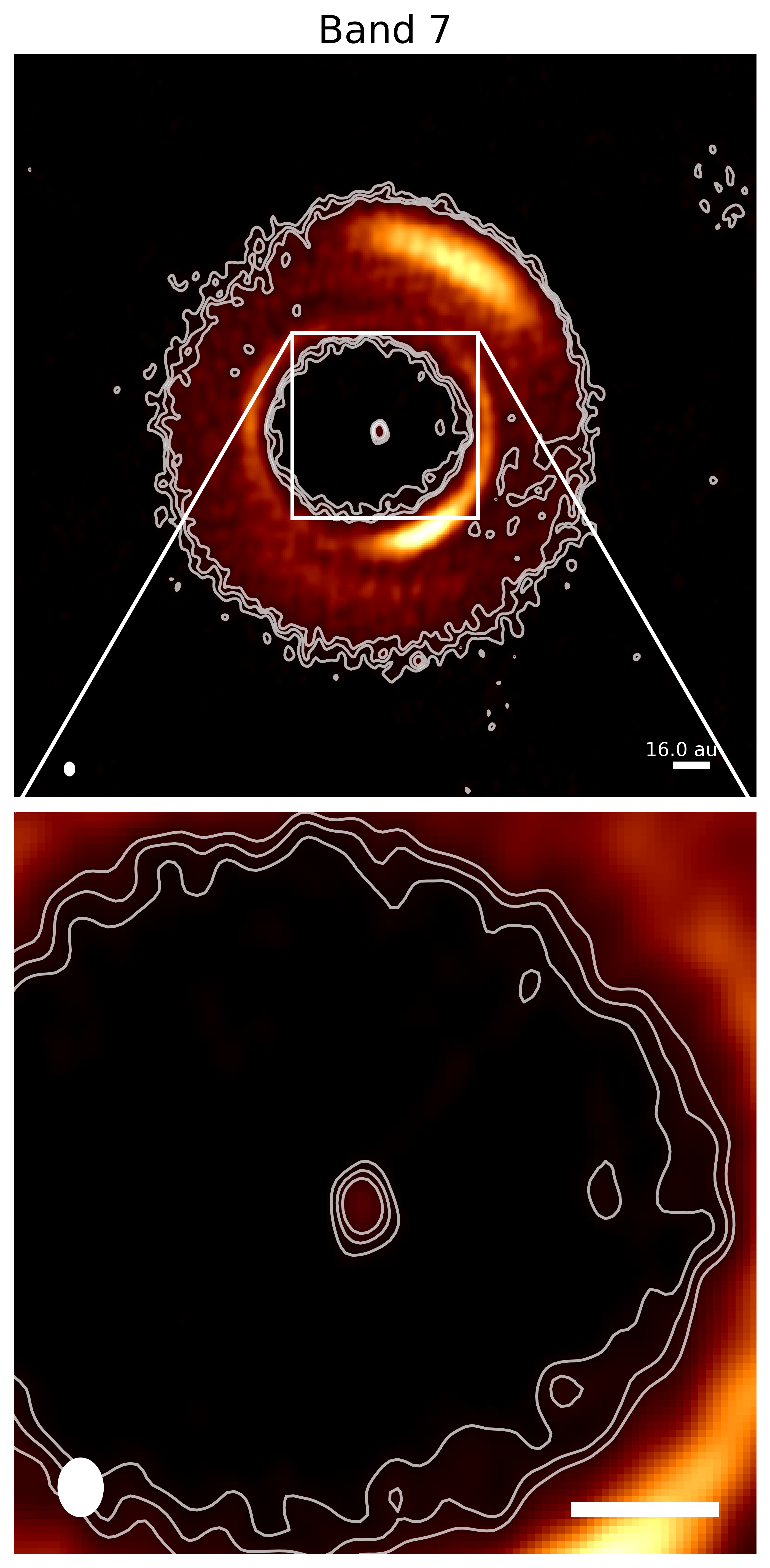}
      \caption{Intensity maps of the disk around MWC758. The image of the disk in Band 7 is show. The images in the first row are $2''.0 \times 2''.0$, while the zooms in the second row are $0''.5 \times 0''.5$.  In each image, the color scale has the peak flux as the maximum, and the image rms as minimum. All bars in the bottom right are $0''.1$ in length, which is $\sim16.0$ au at the distance of the source. White contours show three, five, and seven times the rms of the continuum emission.}
    \label{fig:MWC758}  
\end{figure*}

\begin{figure*} 
   \centering
      \includegraphics[width=0.7\textwidth]{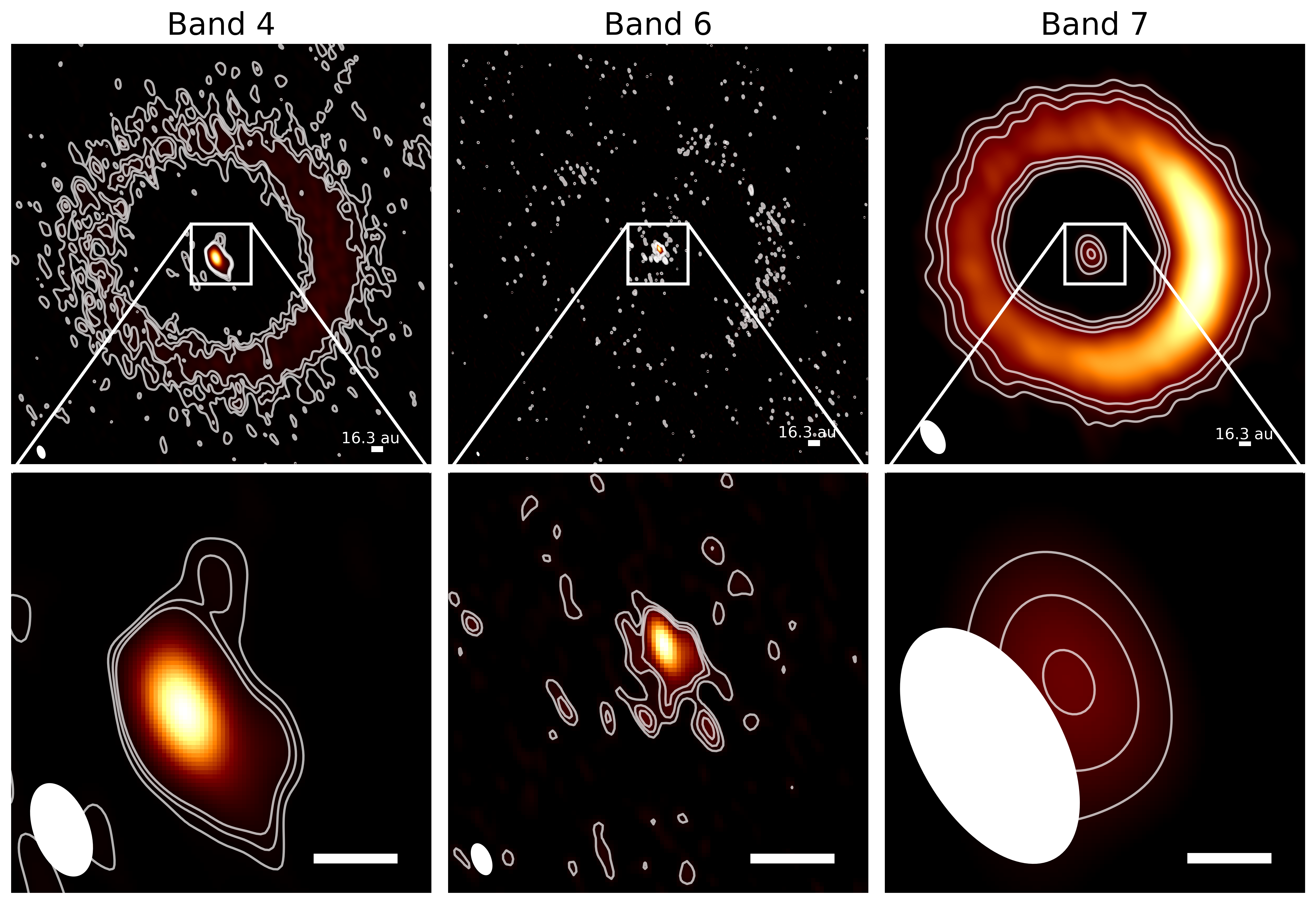}
      \caption{Intensity maps of the disk around AB Aur. The first, second, and third columns show the images of the disk in Band 4, Band 6, and Band 7, respectively. The images in the first row are $3''.5 \times 3''.5$, while the zooms in the second row are $0''.5 \times 0''.5$.  In each image, the color scale has the peak flux as the maximum, and the image rms as minimum. All bars in the bottom right are $0''.1$ in length, which is $\sim16.3$ au at the distance of the source. White contours show three, five, and seven times the rms of the continuum emission.}
    \label{fig:ABaur}  
\end{figure*}

Figures \ref{fig:HD100453}-\ref{fig:ABaur} show the intensity maps of each disk in each analyzed wavelength.


\end{document}